\documentclass[draft,tightenlines,nofootinbib,preprint,aps,eqsecnum,amsmath,amssymb]{revtex4}

\newcommand{\beq}{\begin{equation}}
\newcommand{\eeq}{\end{equation}}
\newcommand{\bea}{\begin{eqnarray}}
\newcommand{\eea}{\end{eqnarray}}

\newcommand{\sgn}{\epsilon}

\begin{document}

\title{On the Transition from the Quantum to the Classical Regime for
Massive Scalar Particles: A Spatiotemporal Approach}

\medskip

\author{Luca Lusanna}

\affiliation{ Sezione INFN di Firenze\\ Polo Scientifico\\ Via Sansone 1\\
50019 Sesto Fiorentino (FI), Italy\\ Phone: 0039-055-4572334\\
FAX: 0039-055-4572364\\ E-mail: lusanna@fi.infn.it}

\author{Massimo Pauri}
\affiliation{Dipartimento di Fisica\\
Universit\'a di Parma, Parco Area Scienze 7/A,\\
43134 Parma, Italy\\
E-mail: pauri@pr.infn.it}

\today

\begin{abstract}

If the classical structure of space-time is assumed to define an
a-priori scenario for the formulation of the structure of quantum
theory (QT), the coordinate representation of the solutions
$\psi(\vec x, t)$ ($\psi({\vec x}_1,..,{\vec x}_N, t)$) of the
Schroedinger equation of a quantum system containing one ($N$)
massive scalar particle has a {\it preferred status}. It is then
possible to perform a multipolar expansion of the density matrix
$\rho(\vec x, t) = |\psi(\vec x, t)|^2$ (and more generally of the
Wigner function) around a space-time trajectory ${\vec x}_c(t)$ to
be properly selected. A special set of solutions $\psi_{EMWF}(\vec
x, t)$, named {\it Ehrenfest monopole wave functions}(EMWF), is
characterized by the conditions that: (i) the quantum expectation
value of the position operator coincides at any time with the
searched classical trajectory, $< \psi_{EMWF} | {\hat {\vec x}} |
\psi_{EMWF}> = {\vec x}_c(t)$, and, (ii) Ehrenfest's theorem holds
for the expectation values of the position and momentum operators.
The first condition implies the vanishing of the 'dipole' term in
the multipolar expansion of the density matrix with respect to such
trajectory. Ehrenfest's theorem applied to EMWF leads then to a {\it
closed Newton equation of motion for the classical trajectory, where
the effective force is the Newton force plus non-Newtonian terms (of
order $\hbar^2$  or higher) depending on the higher multipoles of
the probability distribution $\rho$.} Note that the super-position
of two EMWFs is not an EMWF, a result to be strongly hoped for,
given the possible unwanted implications concerning classical
spatial perception. These results can be extended to N particle
systems and to relativistic quantum mechanics.

Consequently, for the states of a quantum particle which are EMWF,
we get the {\it emergence of a corresponding classical 'effective'
particle following a Newton-like trajectory in space-time}. Note
that this holds true in the standard framework of quantum mechanics,
i.e. by assuming the validity of Born's rule and the individual
system interpretation of the wave function (no ensemble
interpretation). These results are valid without any approximation
(like $\hbar \rightarrow 0$, big quantum numbers,...). Moreover, we
do not commit ourselves to any ontological interpretation of quantum
theory (such as, e.g., the Bohmian one. It will be clear that our
trajectories are not Bohm's trajectories). We will argue that, in
substantial agreement with Bohr's viewpoint, the macroscopic
description of the preparation, certain intermediate steps and the
detection of the final outcome of experiments involving massive
particles are dominated by these 'classical {\it effective}
trajectories'.

This approach can be applied to the point of view of de-coherence
(in which {\it positions} turn out to be selected preferred robust
bases) in the case of a diagonal reduced density matrix $\rho_{red}$
(an {\it improper} mixture) depending on the position variables of a
massive particle and of a pointer. When both the particle and the
pointer wave functions appearing in $\rho_{red}$ are EMWF, the
expectation value of the particle and pointer position variables
becomes a statistical average on a classical ensemble. In these
cases an {\it improper} quantum mixture becomes a {\it classical
statistical} one, thus providing an answer to an open problem of
de-coherence about the emergence of classicality.

Our results cast some light on the so-called {\it problem of the
classical regime} and also provide support to Bohr's point of view,
without adhering to any underlying ontology. Finally, we add some
comments on the possible implications of our results for the theory
of measurement, emphasizing the relevance of the many-body problem
in the description of the macroscopic measuring apparatuses and the
fact that the quantum system under investigation should be
realistically considered as an open quantum subsystem which does not
follow a unitary evolution. With this in view, the 'collapse of the
wave function' with the eigenvalue-eigenvector link and the standard
break of unitarity in the measurement processes would be originated
by an idealized instantaneous approximation of a process localized
in neither space nor time.

The main open point is whether this description of the emergence of
classicality can be extended to experiments with photons, many of
which are in the realm of quantum optics but are described at the
macroscopic level in terms of {\it effective} light-rays of
geometrical optics interpreted as {\it effective} photon
trajectories. Finally, our approach should also be extended to
massive particles with spin.

\end{abstract}

\maketitle

\vfill\eject

\section{Introduction}

In the literature of the last twenty years important new
developments in different domains of the physical research have been
widely accepted as casting some light on the quantum-to-classical
transition. More precisely, they configured the so-called {\it
problem of the classical regime}, i.e. is the question of whether
and how the sweeping success of the classical physical description
(in particular on the macroscopic scale) can be explained in quantum
mechanical terms. There is also agreement on the fact that, even if
the {\it problem of the classical regime} is relatively independent
of the traditional {\it measurement problem}, it is equally
important in assessing the empirical adequacy of quantum theory (QT)
and its interpretations. See Refs.\cite{1,2,3,4,5,6,7} as a sample
of extensive bibliography on these subjects and the associated
issues of the uniqueness of outcomes, of Born's probability rule, of
the acceptance or non-acceptance of the collapse of the wave
function as a real process (in connection with the interpretation of
the wave function itself) and of the eigenvalue-eigenvector link. In
this vast literature, however, the discussion is often characterized
by imprecise, purely verbal statements which do not help to
characterize the conceptual heart of the problem discussed. Finally,
we would like to stress that the essential epistemological
difference between Bohr's traditional viewpoint and the contemporary
emphasis on the {\it problem of the classical regime} lies in the
fact that asserting the latter is tantamount to refusing to take for
granted the classical appearances of our macroscopic world. In
particular, within the so-called de-coherence approach.\medskip

As for the classical limit of quantum theory, at present we have
only non conclusive statements about either the $\hbar \rightarrow
0$ limit (like the WKB approximation) or  the limit of large quantum
numbers. Also, Ehrenfest's standard theorem is essentially a
statement about the expectation value of the position and momentum
operators in a given quantum state: although they show a
quasi-classical behavior, the interpretation of the theorem is still
unclear.\medskip

Non-relativistic quantum mechanics (QM) is used nowadays in low
energy experiments in atomic and molecular physics by taking into
account the electro-magnetic field at the order $1/c$ (note that the
Galilei group cannot be consequently implemented). If one wishes to
take into account relativistic kinematics (i.e., the Poincar\'e
group) one must use relativistic QT (RQT) as an approximation to
relativistic quantum field theory (RQFT) valid below the threshold
of particle production. On the other hand, in particle and nuclear
physics one needs the Fock space of RQFT and its particle
interpretation. This framework is unavoidable when one deals with
photons: in many cases one can exploit at best a quasi-classical
eikonal interpretation with wave fronts replaced by congruences of
light rays (interpreted as trajectories of unspecified 'photons') or
with Gaussian wave-packets of classical light. However, there is no
analogue of Ehrenfest's theorem for the expectation values of the
quantum electro-magnetic field between suitable (coherent?) states.
\medskip

In any actual experiment there is a unique random outcome without
quantum superposition of macroscopically distinguishable states and
with an outcome probability distribution consistent with the Born
rule. The only experimentally relevant quantity is in any case the
density matrix associated with the wave function and its standard
probabilistic interpretation.
\bigskip

It is important to remark that all of the experiments involving
massive particles are planned and interpreted in terms of {\it
'effective' particles (see later) localized up to a small
probability cloud} and adopting a Newtonian (relativistic when
needed) quasi-classical intuition about their motion in space-time.
Actually, in the preparation of any actual experiment (particle
physics at CERN, neutron interferometry, atom interferometry, atoms
in a resonant cavity,...) one produces a beam of {\it effective}
particles following a well defined {\it mean trajectory} leading to
the experimental area, with a well-defined {\it mean value of energy
and momentum }(usually defined by the time-of-flight method). In a
scattering process one deals with incoming beams of this type and
identifies outgoing beams of other {\it effective} particles with
{\it effective} trajectories and 4-momentum. In the outcomes of the
double slit experiment one detects macroscopic traces of {\it
effective} particles on the screen. In atom interferometers the wave
function beyond the beam splitter must be thought of as
approximately localizable in the form of two beams of {\it
effective} atoms in the two arms, {\it notwithstanding the existing
interference effects}. In a radioactive decay an {\it effective}
nucleus localized in a cavity decays and the decay products are
detectable {\it effective} particles. The same happens in the decay
of muons in the atmosphere and in the decay of mesons in EPR
experiments. In quantum optics an {\it effective} atom in a cavity
emits or absorbs photons by interacting with lasers beams, while in
an atomic fountain a beam of {\it effective } atoms feels gravity
and interacts with laser beams. And so on.\medskip

It therefore seems that most of the {\it realizable experiments}
admit a quasi-classical description in terms of {\it effective}
massive particles with a {\it mean trajectory} and a {\it mean value
of 4-momentum}, i.e. {\it effective} entities approximately
localized in space-time. On the other hand, in experiments like the
double slit, in which interference effects dominate, the description
of the propagation of the massive quantum particles from the slits
to the screen is more appropriately provided by using a 'wave'
interpretation. This is the starting point of the {\it pilot wave
description} \cite{8,9} (even considered independently of the
possible addition of hidden extra variables describing, e.g., the
positions of {\it added} 'real' particles as in Bohmian ontological
interpretation). Like in the eikonal approximation of classical
electro-magnetic waves, the wave function is parametrized as a
modulus (the square root of the density matrix) and a phase (an
action variable) while the Schroedinger equation is replaced by two
coupled equations \footnote{It is well-known that they are the {\it
guidance} equation for the velocity field of the tangents to the
rays (implying an equation similar to the Euler equation for a
fluid) and an equation of the {\it Hamilton-Jacobi} type for
classical massive particles.}. Then one looks for the {\it effective
rays} (the world-line of fluid elements named Bohm trajectories)
propagating from the slits to the screen. Avoiding the Bohmian
ontology and exploiting the Feynman path integral representation of
the wave function, such {\it effective rays} could be interpreted as
{\it effective} particles and viewed as an {\it emerging feature} of
the quantum reality.\medskip

Finally, we should ask ourselves whether there could exist some
experimental situations in which neither the 'particle' nor the
'wave' aspect dominates the physical description \footnote{Let us
note that 'particle' and 'wave' seem to be the only possible
spatiotemporal formats to represent a physical entity. In
Ref.\cite{10} a heterodox proposal is advanced concerning the
unavoidable {\it epistemic primacy} of this spatiotemporal
intuition.}: could they be detectable with the existing
experimental technology?
\bigskip

This state of affairs is well reflected in the following passage of
von Weizs\"acker.\medskip

 "We ought not to say 'Every experiment that is even possible {\it
must} be classically described, i.e. {\it localized in space-time}',
but 'Every {\it actual} experiment known {\it to us} "{\it is}"
classically described {\it in this way}, and we do not know how to
proceed otherwise' "(p.128 of Ref.\cite{11}).
\medskip

This citation reflects Bohr's legacy very well (see
Refs.\cite{1,2}), namely the need for some classical distinction
between the quantum system and the apparatus needed by the praxis of
experimentalists. While the measuring instrument may of course be
described as a quantum-mechanical system, it is only possible to
explain its capacity to {\it function as} a measure instrument on
the assumption that the resulting traces of the interaction between
object and instrument can be described in terms of an exchange of
energy and momentum in space and time (i.e. {\it action}). Actually
Bohr stressed:
\medskip

"The unambiguous interpretation of any {\it measurement} must be
essentially framed in terms of classical physical theories, and we
may say that, in this sense, the language of Newton and Maxwell will
remain the language of physics for all time (p.692 of
Ref.\cite{12}). Furthermore:
\medskip

"It is decisive to recognize that, {\it however far the phenomena
transcend the scope of classical physical explanation, the account
of all evidence must be expressed in classical terms}. The argument
is simply that by the word 'experiment' we refer to a situation
where we can tell others what we have done and learned and that,
therefore, the account of the experimental arrangement and of the
results of the observations must be expressed in unambiguous
language with suitable application of the terminology of classical
physics (p. 209 of Ref.\cite{13})."

\medskip

It is well-known that Bohr regarded QT as the universally correct
theory, which would in principle - i.e. given an appropriate
experimental arrangement - imply its application to the description
of macroscopic measurement apparatuses and observers too. However,
Bohr felt that the experimental setup must be described in terms of
classical physics if it is to serve as a measuring instrument at
all, and that there should be a form of separability between the
observed system and the apparatus (the observer). One is then left
with the idea that a shifty quantum-classical divide (the so-called
Heisenberg cut) is a necessary part of the epistemological structure
of QT.

\bigskip

To our knowledge, there has so far been no satisfactory theoretical
support to the description of existing experiments in terms of {\it
effective} quasi-classical particles propagating in space-time.
Moreover, this simulation seems to be independent of the choice of
any solution of the measurement problem and also of the presence of
the environment with its role in de-coherence.

\medskip

In this paper we intend to throw some additional light on the {\it
phenomenology of quantum experimental situations} with its apparent
dominance of what we have called quasi-classical {\it effective}
massive particles. More precisely, the term {\it effective} intends
to stress the nature of the classical structure emerging from the
quantum level by means of a mathematical manipulation which allows
to recover a Newton-like description of such entities. Here we will
not take either the limit $\hbar \rightarrow 0$, like for instance
in the WKB approximation, or the limit of large quantum numbers, or
the $N \rightarrow \infty$ limit for multi-particles states.

\medskip
Lets add a few final remarks on the issue of the so-called {\it
collapse} of the wave function, even if this does not belong to our
main concerns. Actually, a {\it matter-of-fact question} about the
collapse exists only for an 'ontological' interpretation of the wave
function. In that case, however, the very notion of collapse should
be replaced by a dynamic explanation of what goes on during a
measurement process, as is the case, e.g., in the GRW theory
\cite{7} and other attempts to solve the measurement problem (by
means of a unitary, but non-linear, time evolution). If one shares -
as is our case - a {\it weak instrumentalist} viewpoint about QT,
more or less {\it a la Bohr} \cite{14}, and considers the wave
function just as a 'catalogue of probabilities' (as Schroedinger put
it), the issue of collapse becomes essentially a question of words.
Namely, how to describe the information change following the fact
that after a measurement, the 'catalogue of probabilities' must be
redefined. In other words, the physical amplification process that
leads to recording, which is always highly complicated and
irreversible, is such that the 'closure' of a measurement process in
QT, though not yet understood, does not imply any {\it practical}
difficulty. Note, furthermore, that at the experimental level there
is no clear, definitive evidence of the validity of the
eigenvector-eigenvalue link. In conclusion our approach has nothing
to say on the traditional foundational issues of QT, including the
unicity of the outcomes. We shall only offer some suggestions about
the relevance of the effective Newtonian trajectories in identifying
the areas of the detectors where the amplification process, leading
to a final pointer position as a result of a measurement, is
randomly localized.

\bigskip

A collateral discourse should be made concerning the current views
about de-coherence in which the measurement process is {\it
approximately} described by a {\it unitary} transformation involving
the quantum system to be measured, the measuring apparatus and the
environment. In this way, one gets a 'dynamic' explanation of a
sort, of the measuring process in which a collapse phase is not made
explicit and interference tends to be suppressed. Of course, this
last viewpoint is not the right one to be exploited in the cases in
which {\it quantum interference effects} represent just what is to
be made {\it explicit} in a measurement. See, e.g., the insightful
analysis of this point in Refs.\cite{1} (see Sect. 3.3). Our
contribution to the viewpoint of de-coherence will concern only an
instantiation of the important transition from genuinely
non-classical {\it quantum improper mixtures} (mixing 'ignorance and
entanglement') to {\it classical statistical mixture }(ignorance
only), a transformation that becomes almost automatic with our
method, at least in special cases.

\bigskip

This said, we will start from the following assumptions:\medskip

1) the wave function describing a quantum system and satisfying the
standard time-dependent Schroedinger equation encodes all the
probabilistic properties of the given individual system (no ensemble
interpretation);

\medskip

2) space-time structure is presupposed, independently of and before
the technical formulation of QT and its interpretations, even if its
symbolic structure goes - so speak - beyond space-time. The inertial
frames of Galilei and Minkowski space-times define a scenario for
describing matter (here we are not considering fields but particles
only) whose dynamics satisfies either Newtonian and relativistic
classical mechanics (NCM and RCM), or non-relativistic QT and RQT
with the transformation rules connecting inertial frames governed by
the Galilei and Poincar\'e groups, respectively. We shall mainly
discuss the standard non-relativistic case in inertial frames of
Galilei space-time (Newtonian gravity can be present) and we will
extend to the inertial frames of Minkowski space-time subsequently
(but without relativistic gravity). In Ref.\cite{15} there is a
review of the status of understanding of the theory of relativistic
massive either classical or quantum particles and what is known
about extending the results to non-inertial frames (see also
Refs.\cite{16,17}).

\medskip

Starting with this {\it space-time oriented} point of view, a
deterministic point-like classical particle in an inertial frame of
Galilei space-time is described in QT by a wave function solution of
the Schroedinger equation with a given quantum Hamiltonian and
having the full Euclidean 3-space $\Sigma_t$ at time t as
background. It is accordingly assumed that the coordinate
representation has a {\it privileged kinematical and descriptive
status among all the possible bases in Hilbert space}.
\medskip

Given the Schroedinger equation and the space of its solutions, we
exploit Ehrenfest's theorem \cite{18}, which states that the
equation of motion for the expectation value of the position and
momentum operators have {\it some resemblance} with Newton-Hamilton
equations of motion. If Hilbert space is a suitable function space,
the recent mathematical developments of Ref.\cite{19} allow to prove
Ehrenfest's theorem for a system of N particles with mutual Coulomb
interaction (i.e. for the main system behind molecular physics).
Moreover, both the wave function and the density matrix of a quantum
system have the whole 3-space at a fixed time as a support, even
when they are strongly peaked like the Gaussian wave packets with
minimal position and momentum indeterminacy. Furthermore, we note
that even if, in general, Ehrenfest expectation values for well
peaked wave functions have non-classical behavior on long times, it
has been found that in many cases the spread of wave packets stops
and leads to a revival of the original quasi-classical behavior
\cite{20}. Finally, let us stress that the crucial obstacle for
exploiting Ehrenfest's theorem is due to the fact that the {\it
time-dependent mean values of position operators are not
trajectories of classical Newton (or relativistic) particles}
\cite{4}.

\bigskip

The technical input to face {\it the problem of the classical
regime} comes from the remark that in classical physics (matter or
fluids interacting with the electro-magnetic and/or gravitational
field) nobody is able to describe extended objects: actually they
are always described by making a {\it multi-polar expansion} of
their energy-momentum tensor (when known) {\it around a classical
world-line} together with postulated  equations of motion for the
multi-poles. See Ref.\cite{21} for a review given by one of the
fathers of this method and for the analiticity properties required
for the validity of the multi-polar expansion. See also
Ref.\cite{22} for applications in special relativity.

\medskip

We will accordingly study the multipolar expansion around a given
space-time trajectory of the density matrix $\rho(\vec x,t) =
\psi^*(\vec x, t)\, \psi(\vec x, t)$ (and of other bilinears like $
- \psi^*(\vec x, t)\, i\hbar\, {{\partial}\over {\partial\, x^r}}\,
\psi(\vec x, t)$, needed for the expectation value of the
momentum).\medskip

We then search for a special set of solutions $\psi_{EMWF}(\vec x,
t)$ of the Schroedinger equation for the quantum system with a given
Hamiltonian, to be named {\it Ehrenfest monopole wave functions}
(EMWF), having the following properties:\medskip

a) {\it The quantum expectation value of the position operator of
the massive particle, at any time, coincides with the searched
classical space-time trajectory},  $< \psi_{EMWF} | {\hat {\vec x}}
| \psi_{EMWF} > = {\vec x}_c(t)$, (the {\it 'monopole' as a
classical point-like effective massive particle}). This implies the
vanishing of the 'dipole' of the density matrix with respect to the
trajectory. Note that the classical 'coherent states' satisfy the
dipole condition.
\medskip

b) Due to the first half of Ehrenfest equations, the expectation
value of the momentum operator gives the classical momentum.

\medskip

c) The second half of Ehrenfest equations provides a {\it
deterministic Newtonian equation of motion} for the space-time
trajectory, where the {\it effective} force is the Newton force plus
non-Newtonian terms (of order $\hbar^2$ or higher) depending upon
higher multi-poles of the probability distribution $\rho$ which
characterize the probabilistic properties of the quantum entity.

\medskip

d) The emergent classical space-time trajectories will clearly
depend on the chosen EMWF of the given quantum system. Different
EMWFs, e.g. eigenstates of non-commuting operators, will have
different associated classical trajectories reflecting the
properties of the observables associated to such operators. If, in
particular, two EMWFs are eigenfunctions corresponding to different
eigenvalues of the same operator so that their associated classical
trajectories carry information about the properties of the
observables of such operator, all of the superpositions of such wave
functions which are {\it also} EMWFs will still have classical
trajectories unrelated to these properties. They will describe, so
to speak, different and non-intrinsic or state dependent,
properties. One could even say that our emergent Newtonian world is
{\it contextual}. Finally note that, as is highly desirable, the
notion of superposition of classical trajectories is structurally
meaningless.

\medskip

Finally, the multipolar expansion of the associated Wigner function
$W(\vec x, \vec p, t) = \int d^3\xi\, e^{- {i\over {\hbar}}\, \vec p
\cdot \vec \xi}\, \psi^*(\vec x - {1\over 2}\, \vec \xi, t)\,
\psi(\vec x + {1\over 2}\, \vec \xi, t)$ allows to study the
expectation value of any operator.

\bigskip

It is important to stress that our method turns out to be applicable
to N particle systems too: the density matrix associated to a wave
function $\psi({\vec x}_1,..., {\vec x}_N, t)$, living at each
instant $t$ in a 3N-dimensional configuration space, gives rise to a
set of N effective trajectories inside the Euclidean 3-space when
{\it all} of the dipolar moments of its multiple multipolar
expansion vanish around these trajectories. This is a natural
solution of a long standing interpretational problem of the
N-particle wave functions.

\bigskip

Unlike all other non-relativistic treatments which have problems
with their relativistic extension, our approach to the {\it problem
of classical regime} works exactly in the same way within the new
consistent RQM developed in Ref.\cite{16}. The main feature is a new
kind of {\it non-locality}\footnote{Having nothing to do with
standard quantum non-locality.} and {\it spatial non-separability}
of the description of N-particle system, induced by the Lorentz
signature of space-time. It requires a clock synchronization
convention for the definition of instantaneous 3-spaces (reduction
of the problem of relative times among the particles), and entails
non-local definitions of the collective variables (as is well known,
there is no unique notion of center of mass in relativistic theory).
Consequently, a consistent description of the N particles can be
obtained only in terms of {\it relative 3-coordinates} (spatial
non-separability of subsystems) and entails - unlike the Galilean
case - a dependence of the Lorentz boosts upon the inter-particle
interactions. Then, the Hilbert space of a relativistic composite
system is {\it not} the tensor product of the Hilbert spaces of the
subsystems. It can only be defined as the tensor product of the
Hilbert space of the decoupled non-covariant (non-local, i.e. non
measurable) center of mass times the Hilbert space of {\it relative
motions}.

\bigskip

Another relevant feature of our approach is its applicability within
the framework of de-coherence. Namely, given a reduced density
matrix ({\it improper quantum mixture} based on both ignorance and
entanglement) in the selected preferred robust positional basis for
both a massive particle and the pointer of a macroscopic apparatus,
we shall show that, if all the particle and pointer wave functions
appearing in this description are EMWF, the {\it improper quantum
mixture} can be replaced by a {\it classical statistical proper
mixture} based only on our ignorance about the localization of the
particle and the pointer effective trajectories in the region
allowed by the experimental apparatus. This should contribute to a
partial clarification of one of the main open problems of
de-coherence.

\bigskip

Some concluding remarks are advanced on experiments involving
photons instead of massive particles by taking into account recent
{\it weak} measurements of the average trajectory of single photons
in the double slit experiment \cite{25}. The adaptation of our
approach to the non trivial cases of Pauli and Dirac spinors will be
dealt with in a separate paper.

\bigskip

While Section II is dedicated to the main technical results, in
Section III we give the multipolar expansion of the Wigner function.
The results are extended to two particle systems in Section IV.

In Section V we study relativistic quantum particles in the inertial
rest frame.

Section VI is devoted to a comparison of experiments in which the
'particle' aspects of the wave function dominate, versus cases in
which the 'wave' aspects dominate.

In Section VII we review the support given by de-coherence to the
preferred status of position measurements and we show how EMWFs
suggest a natural transition from quantum improper mixtures to
classical statistical ones.

A concluding Section contains some general remarks on the problem of
classical regime, its extension to the relativistic level and the
potentialities of this approach for foundational problems of QM.

In Appendix A we review the existing mathematical problems
concerning spatial localization.

\vfill\eject

\section{The Monopole Ehrenfest Wave Functions}

Let us consider a quantum point-particle in non-relativistic QT
whose normalized wave function $\psi(\vec x, t) =  < \, \vec x |\,
\psi(t) >\, \in\, L^2(R^3)$  is a solution of the Schroedinger
equation $i\hbar {{\partial}\over {\partial t}}\, \psi(\vec x, t) =
\hat H\, \psi(\vec x, t)$ with classical Hamiltonian $H = {{{\vec
p}^2}\over {2m}} + V(\vec x)$ \footnote{If $A$ is a classical
quantity, $\hat A$ denotes the associated quantum operator.}. We
work in the coordinate representation, where ${\hat {\vec x}} = \vec
x$ and ${\hat {\vec p}} = - i\hbar\, {{\partial}\over {\partial \vec
x}}$ and where the density matrix operator $\hat \rho (t) = |\psi(t)
>\, < \psi(t) |$ takes the following form (the second line is
implied by the Schroedinger equation)

\bea
 \rho_{(o)}(\vec x, t) &{\buildrel {def}\over =}&  < \vec x | \hat
 \rho(t) | \vec x > = |\psi(\vec x, t)|^2,\qquad \int d^3x\,
 \rho_{(o)}(\vec x, t) = 1,\nonumber \\
 &&{}\nonumber \\
 \partial_t\, \rho_{(o)}(\vec x, t) &=& \vec \partial \cdot
 \Big(\psi^*\, {{{\hat {\vec p}}}\over {2m}}\, \psi - ({{{\hat {\vec
 p}}}\over {2m}}\, \psi^*)\, \psi\Big)(\vec x, t)\, {\buildrel
 {def}\over =}\, - {1\over m}\, \vec
 \partial \cdot {\vec \rho}_{(1)}(\vec x, t),\nonumber \\
   && {\vec \rho}_{(1)}(\vec x, t) = -
 {{i\hbar}\over 2}\,
 \Big(\psi^*\, {{\partial\, \psi}\over {\partial\, \vec x}} - {{\partial\,
 \psi^*}\over {\partial\, \vec x}}\, \psi\Big)(\vec x, t).
 \label{2.1}
 \eea

The expectation values of the position ${\hat {\vec x}} = \vec x$
and momentum ${\hat {\vec p}} = - i\hbar\, {{\partial}\over
{\partial \vec x}}$ operators in the state $\psi(\vec x, t)$
are\medskip

\beq
 < {\hat {\vec x}} >_{\psi(t)}\, = \int d^3x\, \vec x\,
 \rho_{(o)}(\vec x, t),\qquad < {\hat {\vec p}} >_{\psi(t)}\, =
 - i\hbar\, \int d^3x \Big(\psi^*\, {{\partial\, \psi}\over
 {\partial\, \vec x}}\Big)(\vec x, t)  = \int
 d^3x\, {\vec \rho}_{(1)}(\vec x, t),
 \label{2.2}
 \eeq

 \noindent and the Ehrenfest theorem implies

 \beq
 {d\over {dt}}\,  < {\hat {\vec x}} >_{\psi(t)}\, = {1\over m}\,
 < {\hat {\vec p}} >_{\psi(t)},\qquad
 {d\over {dt}}\, < {\hat {\vec p}} >_{\psi(t)}\, = < - {{\partial\,
 V(\vec x)}\over {\partial\, \vec x}} >_{\psi(t)}.
 \label{2.3}
 \eeq

\medskip

Then, following Refs.\cite{5,6}, we perform a multipolar expansion
of the distribution functions $\rho_{(o)}(\vec x, t)$ and ${\vec
\rho}_{(1)}(\vec x, t)$  around a given classical trajectory ${\vec
x}_c(t)$  in Galilei space-time

\bea
 \rho_{(o)}(\vec x, t) &=& \rho_{(o) o}({\vec x}_c(t), t)\, \delta^3(\vec x - {\vec x}_c(t))
 + \sum_{n=1}^{\infty}\, {{(-)^n}\over {n!}} \sum_{r_1,..,r_n}\,
 \rho_{(o) n}^{r_1..r_n}({\vec x}_c(t), t)\, {{\partial^n\, \delta^3(\vec x -
 {\vec x}_c(t))}\over {\partial\, x^{r_1}\, ...\, \partial\, x^{r_n}}},\nonumber \\
 &&{}\nonumber \\
 &&\rho_{(o) n}^{r_1..r_n}({\vec x}_c(t), t)\, =\, \int d^3x\, \Big(x^{r_1} - x_c^{r_1}(t)\Big)
 ... \Big(x^{r_n} - x_c^{r_n}(t)\Big)\, \rho_{(o)}(\vec x, t),\nonumber \\
 &&\rho_{(o) o}({\vec x}_c(t), t)\, =\, \int d^3x\, \rho_{(o)}(\vec x, t) =
 \rho_{(o)o}(t) = 1,\nonumber \\
 &&{}
 \label{2.4}
 \eea

\bea
 {\vec \rho}_{(1)}(\vec x, t) &=& {\vec \rho}_{(1)o}({\vec x}_c(t),
 t)\, \delta^3(\vec x - {\vec x}_c(t))
 + \sum_{n=1}^{\infty}\, {{(-)^n}\over {n!}} \sum_{r_1,..,r_n}\,
 {\vec \rho}{}_{(1)n}^{r_1..r_n}({\vec x}_c(t), t)\, {{\partial^n\, \delta^3(\vec x -
 {\vec x}_c(t))}\over {\partial\, x^{r_1}\, ...\, \partial\, x^{r_n}}},\nonumber \\
 &&{}\nonumber \\
 &&{\vec \rho}{}_{(1)n}^{r_1..r_n}({\vec x}_c(t), t)\,
 =\, \int d^3x\, \Big(x^{r_1} - x_c^{r_1}(t)\Big)
 ... \Big(x^{r_n} - x_c^{r_n}(t)\Big)\, {\vec \rho}_{(1)}(\vec x, t) =\nonumber \\
 &=& {{i\hbar}\over 2}\, \int d^3x\, \Big(x^{r_1} - x_c^{r_1}(t)\Big)
 ... \Big(x^{r_n} - x_c^{r_n}(t)\Big)\,
 \Big(\psi^*\, {{\partial\, \psi}\over {\partial\, \vec x}} - {{\partial\,
 \psi^*}\over {\partial\, \vec x}}\, \psi\Big)(\vec x, t).\nonumber \\
 &&{}
 \label{2.5}
 \eea

\medskip

If the density $|\psi(\vec x, t)|^2$ is concentrated in a spatial
region of the order of the de Broglie wavelength, the classical
trajectory ${\vec x}_c(t)$ should be located within this region.
Accordingly we shall adopt the notation  ${\vec x}_c(t) = {\vec
x}_{\psi}(t)$.\medskip

The higher multipoles  are symmetrical in the indices $r_1,..,r_n$:
$\rho_{(o)n}^{r_1..r_n}({\vec x}_{\psi}(t), t) =
\rho_{(1)n}^{(r_1..r_n)}({\vec x}_{\psi}(t), t)$ and ${\vec
\rho}{}_{(1)n}^{r_1..r_n}({\vec x}_{\psi}(t), t) = {\vec
\rho}{}_{(1)n}^{(r_1..r_n)}({\vec x}_{\psi}(t), t)$. The multipoles
$\rho_{(o)n}^{r_1..r_n}({\vec x}_{\psi}(t), t)$ are of order
$\hbar^n$ when their spatial extent is concentrated in such a
region; otherwise their order is larger than $\hbar^n$.

\bigskip

By using Eqs.(\ref{2.4}) and (\ref{2.5}), the expectations values of
position and momentum operators become

\bea
 < {\hat {\vec x}} >_{\psi(t)}\, &=& \int d^3x\, \vec x\, \rho_{(o)}(\vec x, t)
 = {\vec x}_{\psi}(t) + {\vec \rho}_{(o) 1}({\vec x}_{\psi}(t), t),\nonumber \\
 &&{}\nonumber \\
 < {\hat {\vec p}} >_{\psi(t)}\, &=& \int d^3x\, {\vec
 \rho}_{(1)}(\vec x, t) = {\vec \rho}_{(1)o}({\vec x}_{\psi}(t), t)
 {\buildrel {def}\over =} {\vec \rho}_{(1)o}(t),
 \label{2.6}
 \eea

\medskip

\noindent where  ${\vec \rho}_{(o) 1}({\vec x}_{\psi}(t), t)$ is the
\emph{dipole} term of the distribution $\rho_{(o)}(\vec x,
t)$.\bigskip

Note that {\it the trajectory still needs to be specified}. We know,
for example, that when the multipolar expansion is used to describe
extended objects like the Earth, the trajectory is fixed by
requiring that it be coincident with some {\it relevant collective
variable} (as the center of mass in non-relativistic physics). Since
our aim here is to extract some classical features from the quantum
probability distribution $|\psi(\vec x, t)|^2$, it seems natural
{\it impose} that the quantum expectation value of the {\it position
operator} coincides for any t with the corresponding point ${\vec
x}_{\psi}(t)$ of the classical spatial trajectory to be chosen.
However, due to the first of Eqs.(\ref{2.6}), this is only possible
for those wave functions whose density matrix has a \emph{vanishing
dipole} with respect to the very trajectory itself, i.e., if

\bea
 < {\hat {\vec x}} >_{\psi(t)}\, &=& {\vec x}_{\psi}(t),\nonumber \\
 &&{}\nonumber \\
 \Rightarrow&&   {\vec \rho}_{(o) 1}({\vec x}_{\psi}(t), t) = \int d^3y\,
 \vec y\, \rho_{(o)}({\vec x}_{\psi}(t) + \vec y, t) = 0,
 \label{2.7}
 \eea

 \medskip

The wave functions satisfying Eq.(\ref{2.7}) will be named {\it null
dipole wave functions} (NDWF).\medskip

Due to Eq.(\ref{2.7}), the dipole certainly vanishes for all the
wave functions that are eigenfunctions of the parity operator (a
conserved quantity in QT): $\psi_{\pm}(- \vec x, t) = \pm
\psi_{\pm}(\vec x, t)$ if we choose the time axis ${\vec
x}_{\psi}(t) = 0$  as the classical trajectory. This case
corresponds to an isolated quantum system 'at rest' in the origin
(e.g, an atom in a cavity). For instance one can build a complete
set of NDWFs by considering a basis of energy eigenfunctions of the
given quantum system with definite parity (any other basis of
eigenfunctions of a maximal set of commuting operators including
parity will contain NDWFs only). Any other wave function, including
the NDWFs, can be decomposed on this basis, but the multipolar
expansion of the NDWF ones will identify trajectories different from
those associated to the basis.
\medskip

\medskip

Another class of wave functions with vanishing dipole are the ones
whose density matrix has a spatial distribution such that for any
{\it t}, it holds $\rho_{(o)}({\vec x}_{\psi}(t) - \vec y, t) =
\rho_{(o)}({\vec x}_{\psi}(t) + \vec y, t)$. Note that wave
functions having this symmetry exist for every quantum system
describing e.g. collimated beams of massive particles. \medskip

Finally, more general wave functions should possibly exist,
satisfying Eq.(\ref{2.7}), without any explicit symmetry at all.

\bigskip

Let us stress, however, that {\it not all of the NDWF satisfy the
Ehrenfest theorem}. A simple counterexample is the following:
consider an energy wave function $e^{- i\, E_n\, t}\, \psi_n(\vec
x)$ of a given quantum system, which is also a parity eigenfunction
with $\psi_n(- \vec x) = \pm\, \psi_n(\vec x)$. This is a NDWF with
a vanishing expectation value of the position operator $< {\hat
{\vec x}} >_{\psi_n} = 0$ but with a non-vanishing time-independent
expectation value $< {\hat {\vec p}} >_{\psi_n}$ of the momentum
operator. Consequently, Eqs.(\ref{2.3}) cannot hold for them.
\medskip

We shall then call {\it Ehrenfest monopole wave functions} (EMWF)
those NDWFs for which both Eqs.(\ref{2.3}) hold. The important issue
here being the implications of the Ehrenfest theorem for the
associated classical trajectories ${\vec x}_{\psi}(t)$.

\bigskip

For EMWF the first half of Eqs.(\ref{2.3}) implies  that {\it the
expectation value (\ref{2.6}) of the momentum operator coincides
with the expression of the  momentum of the classical trajectory}.
Actually

\beq
 < {\hat {\vec p}} >_{\psi(t)} = {\vec \rho}_{(1)o}({\vec x}_{\psi}(t), t)
 = {\vec \rho}_{(1)o}(t) =
  m\, {\dot {\vec x}}_{\psi}(t)\, {\buildrel {def}\over =} {\vec p}_{\psi}(t).
 \label{2.8}
 \eeq

Here ${\vec p}_{\psi}(t)$ is the classical momentum of a Newtonian
particle, which is a quantity that can be measured by means of,
e.g., the time-of-flight method.

 \medskip

Then, the second half of Ehrenfest equations (\ref{2.3}) implies

\bea
 m\, {\ddot {\vec x}}_{\psi}(t) &=& < - {{\partial\,
 V(\vec x)}\over {\partial\, \vec x}} >_{\psi(t)}
 = {\vec F}_{\psi}(t) = - {{\partial\,
 V({\vec x}_{\psi}(t))}\over {\partial\, {\vec x}_{\psi}}} -\nonumber \\
 &-& \sum_{n=2}^{\infty}\, {1\over {n!}}\, \sum_{r_1..r_n}\,
 \rho_{(o) n}^{r_1..r_n}({\vec x}_{\psi}(t), t)\, {{\partial^{n+1}\, V({\vec x}_{\psi}(t))}
 \over {\partial\, {\vec x}_{\psi}\, \partial\, x_{\psi}^{r_1}\, ... \partial\,
 x_{\psi}^{r_n}}},
 \label{2.9}
 \eea

\noindent namely a {\it deterministic} Newton-like equation, in
which the standard Newton force is complemented with {\it
$\hbar$-dependent non-Newtonian forces} determined by the {\it
higher multi-poles} (of order $\hbar^2$ or higher) of the
probability distribution function $\rho_{(o)}(\vec x, t)$. Note that
these higher multi-poles play the role of a 'quantum guidance' of
the classical {\it effective} trajectory, although in a sense
different from the case of Bohmian mechanics. Finally, the
quadrupole term $\rho^{r_1r_2}_{(o)2}$ carries the information on
the standard deviation $\triangle\, x^r$ of the $\vec
x$-distribution, appearing in the {\it uncertainty relations} (see
after, Eq.(\ref{3.13})).
\medskip

The Cauchy data at the initial time $t_o$ for these Newton-like
equations of motion are ${\vec x}_{\psi}(t_o) = < {\hat {\vec x}}
>_{\psi(t_o)}$ and ${\dot {\vec x}}_{\psi}(t_o) = {1\over m}\,
< {\hat {\vec p}} >_{\psi(t_o)}$, i.e. the mean value of position
and momentum operator at the initial time.
\medskip

In conclusion, the  EMWF share the following dual nature:\medskip

a) on the one hand, they are probability waves describing the
standard QT probability distribution for the localization of the
particle;\medskip

b) on the other hand, they also characterize a deterministic
Newton-like trajectory in Galilei space-time with deviations of
order $O(\hbar^2)$ from a true classical Newtonian trajectory. It
seems that in such cases the efficacy of the Newtonian intuition of
experimentalists in visualizing atoms and molecules is strongly
supported.\medskip

Furthermore, if $\psi_1(\vec x, t)$ and $\psi_2(\vec x, t)$ are two
generic EMWFs for a given system with 'classical' trajectories
${\vec x}_{\psi_1}(t)$ and ${\vec x}_{\psi_2}(t)$ respectively,
their linear superposition $\psi_{12}(\vec x, t) = \alpha\,
\psi_1(\vec x, t) + \beta\, \psi_2(\vec x, t)$ may or may not be an
EMWF. In the case that it is an EMWF centered on a trajectory ${\vec
x}_{\psi_{12}}(t)$  we have

\bea
  < {\hat {\vec x}} >_{\psi_{12}(t)} &=& {\vec x}_{\psi_{12}}(t)
  + {\vec \rho}_{(o)1}({\vec x}_{\psi_{12}}(t), t) =\nonumber \\
  &=& {\vec x}_{\psi_1}(t) + {\vec x}_{\psi_2}(t) +
 \int d^3x\, \vec x\, \Big(\alpha^*\, \beta\, \psi_1^*\, \psi_2 + \alpha\,
 \beta^*\, \psi_2^*\, \psi_1\Big)(\vec x, t) \not=\nonumber \\
 &\not=& {\vec x}_{\psi_1}(t) + {\vec x}_{\psi_2}(t).
 \label{2.10}
 \eea

The interference term is real but not definite positive and does not
admit a multipolar expansion, because it was defined only for the
density function $|\psi(\vec x, t)|^2$. Consequently, the
superposition principle cannot hold in general in the set of EMWFs,
and superposition of classical trajectories is, as desirable, a
meaningless notion. This is not in contradiction with the
possibility of expressing an EMWF on a complete basis of NDWFs with
definite parity. While NDWFs not in the EMWF class do not have
classical trajectories satisfying Newton-like equations of motion,
an EMWF superposition of NDWFs will have a classical trajectory
satisfying Newton-like equations of motion. Coherent states are a
relevant example of EMWFs.

\medskip

Note, therefore, that the existence of the multipolar expansion
around a well defined classical trajectory, identifying NDWF, and
the satisfaction of the Ehrenfest theorem, identifying EMWF with
Newton-like classical trajectories, express two distinct
mathematical properties.

\bigskip

The expectation value of the Hamiltonian operator between NDWF (not
EMWF) which are also energy eigenfunctions, i.e. $\Psi_n(\vec x, t)
= e^{{i\over {\hbar}}\, E_n\, t}\, \psi_n(\vec x)$, $\hat H\,
\psi_n(\vec x) = E_n\, \psi_n(\vec x)$, is discussed at the end of
next Section.

\medskip

Let us consider instead the expectation value of the angular
momentum operator ${\hat L}^i = {1\over 2}\, \epsilon^{ijk}\, {\hat
x}^j\, {\hat p}^k$ in a state EMWF. By using Eqs.(\ref{2.5}) and
(\ref{2.8}) we get

\bea
 < {\hat L}^i >_{\psi(t)} &=&  \epsilon^{ijk}\, \int
 d^3x\, x^j\, \rho_{(1)}^k(\vec x, t) =\nonumber \\
 &=&  \epsilon^{ijk}\, \Big[x^j_{\psi}(t)\, \rho^k_{(1)o}({\vec x}_{\psi}(t), t)
 + \rho_{(1)1}^{k\, j}({\vec x}_{\psi}(t), t)\Big] =\nonumber \\
 &=& L^i_{\psi} +  \epsilon^{ijk}\, \rho_{(1)1}^{k\, j}({\vec x}_{\psi}(t),
 t),
 \label{2.11}
 \eea

\noindent where $L^i_{\psi} = \epsilon^{ijk}\, x^j_{\psi}(t)\, m\,
{\dot x}^k_{\psi}(t)$ is the angular momentum of the deterministic
Newtonian particle. Therefore we have $< {\hat L}^i
>_{\psi(t)} = L^i_{\psi}$ only for those EMWFs such that the
antisymmetric part of the dipole $\rho_{(1)1}^{k\, j}({\vec
x}_{\psi}(t), t)$ of the distribution function ${\vec
\rho}_{(1)}(\vec x, t)$ vanishes. In this case, if the EMWF is  a
normalizable eigenfunction $\psi_{lm}(\vec x, t)$ of ${\hat {\vec
L}}^2$ and of ${\hat L}^3$ with eigenvalues $l$ and $m$ (${\hat
{\vec L}}^2\, \psi_{lm} = \hbar\, l\, (l + 1)\, \psi_{lm}$, ${\hat
L}^3\, \psi_{lm} = \hbar\, m\, \psi_{lm}$), from Eq.(\ref{2.8}) we
get $\hbar\, m = < {\hat L}^3 >_{\psi_{lm}(t)} = x^1_{\psi_{lm}}\,
\rho^2_{(1)o}(t) - x^2_{\psi_{lm}}\, \rho^1_{(1)o}(t) =
L^3_{\psi_{lm}}$. With the higher distribution functions of the next
Section one could see which higher multipoles contribute to $\hbar\,
l\, (l + 1) = < {\hat {\vec L}}^2 >_{\psi_{lm}(t)}$ and which
connection may be established with ${\vec L}^2_{\psi_{lm}}$.

\bigskip

Let us remark that in many-particle systems with mutual interactions
one must perform the separation of variables between the free
decoupled center of mass and relative motions: an Ehrenfest theorem
holds true for such relative variables only, but not for the center
of mass if it is in a momentum eigenstate. Indeed for a plane wave
function $\psi(\vec x, t) = e^{i\, (E\, t - \vec p \cdot \vec x)}$,
which  is not normalizable, we get $\rho(\vec x, t) = 1$, $< {\hat
{\vec x}} >_{\psi} = 0$, $< {\hat {\vec p}} >_{\psi} = \vec p$.
Therefore the Ehrenfest theorem does not hold and there is no
semi-classical visualization for this improper description of a free
particle (commonly used for the center-of-mass free motion in
scattering theory). This is not true for realistic scattering
events, in which there are Gaussian-like wave packets of effective
particles collimated towards the scattering region: they are not
eigenfunctions of the momentum operator and the effective mean
momentum of the beams is evaluated with time-of-flight methods (so
that strictly speaking it is not a momentum eigenvalue).

\medskip

Also the improper non-normalizable localized states $\psi(\vec x, t)
= \delta^3(\vec x - {\vec x}_o)$ have no semi-classical
visualization \footnote{See Section 4.1 of Ref.\cite{1} for
discretized position measurements: obviously, only localization in
finite intervals can be described with normalizable wave
functions.}.

\medskip

Moreover, see Appendix A for comments about the mathematical
problems concerning spatial localization.

\vfill\eject

\section{The Multipolar Expansion of the Wigner function}

As shown in Refs. \cite{26}, \cite{4},  the expectation value of an
operator  $\hat \Omega({\hat {\vec x}}, {\hat {\vec p}})$ function
of the position and momentum operators, in the state $\rho(\vec x,
t)$, can be written

\beq
 < \hat \Omega({\hat {\vec x}}, {\hat {\vec p}}) >_{\psi(t)} =
 \int {{d^3x\, d^3p}\over {(2\pi\, \hbar)^3}}\, \Omega_W(\vec x, \vec p)\,
 W(\vec x, \vec p, t).
 \label{3.1}
 \eeq

\noindent This  expression depends on the following two
quantities:\medskip

i) the Weyl symbol of the operator $\hat \Omega({\hat {\vec x}},
{\hat {\vec p}})$, which  is defined as (see the second of
Refs.\cite{16} for more details)

\beq
 \Omega_W(\vec x, \vec p) = \int d^3\xi\, e^{{i\over {\hbar}}\,
 \vec p \cdot \vec \xi}\, < \vec x - {1\over 2}\, \vec \xi |\,
 \hat \Omega({\hat {\vec x}}, {\hat {\vec p}})\, | \vec x + {1\over
 2}\, \vec \xi >;
 \label{3.2}
 \eeq

\noindent the Weyl symbols of the position and momentum operators
are ${\vec x}_W(\vec x, \vec p) = \vec x$ and ${\vec p}_W(\vec x,
\vec p) = \vec p$;\medskip

ii) the Wigner function (a non-definite-positive distribution
function on phase space; it is definite-positive only for Gaussian
coherent states)

\bea
 W(\vec x, \vec p, t) &=& \int d^3\xi\, e^{{i\over {\hbar}}\, \vec p \cdot \vec
 \xi}\, < \vec x - {1\over 2}\, \vec \xi |\, \hat \rho(t)\, | \vec x
 + {1\over 2}\, \vec \xi > =\nonumber \\
 &=& \int d^3\xi\, e^{{i\over {\hbar}}\, \vec p \cdot \vec
 \xi}\, \psi(\vec x - {1\over 2}\, \vec \xi, t)\, \psi^*(\vec x + {1\over 2}\,
 \vec \xi, t).
 \label{3.3}
 \eea

\bigskip

Let us make the following Cauchy expansions

\bea
 \psi(\vec x - {1\over 2}\, \vec \xi, t) &=& \psi(\vec x, t) +
 \sum_{a=1}^{\infty}\, {{(-)^a}\over {2^a}}\, \sum_{a_1\, .. a_a}\,
 \xi^{a_1}\, .. \xi^{a_a}\, {{\partial^a\, \psi(\vec x, t)}\over
 {\partial\, x^{a_1}\, .. \partial\, x^{a_a}}},\nonumber \\
 \psi^*(\vec x + {1\over 2}\, \vec \xi, t) &=& \psi^*(\vec x, t) +
 \sum_{b=1}^{\infty}\, {1\over {2^b}}\, \sum_{b_1\, .. b_b}\,
 \xi^{b_1}\, .. \xi^{b_b}\, {{\partial^b\, \psi^*(\vec x, t)}\over
 {\partial\, x^{b_1}\, .. \partial\, x^{b_b}}}.
 \label{3.4}
 \eea

Then, Eq.(\ref{3.1}) becomes ($c = a + b$)

\bea
 W(\vec x, \vec p, t) &=&\int d^3\xi\, e^{{i\over {\hbar}}\, \vec p \cdot \vec
 \xi}\, \sum_{a=0}^{\infty}\,
 \sum_{b=0}^{\infty}\, {{(-)^a}\over {2^{a+b}}}\, \sum_{d_1.. d_{a+b}}\, \xi^{d_1+ .. + d_a
 + d_{a+1} +.. +d_{a+b}}\nonumber \\
 && {{\partial^b\, \psi^*(\vec x, t)}\over {\partial\, x^{d_{a+1}}\,
 ... \partial\, x^{d_{a+b}}}}\, {{\partial^a\, \psi(\vec x, t)}\over {\partial\,
 x^{d_1}\, ... \partial\, x^{d_a}}} =\nonumber \\
 &=&\int d^3\xi\, e^{{i\over {\hbar}}\, \vec p \cdot \vec
 \xi}\, \sum_{a=0}^{\infty}\, (-)^a\, \sum_{c=a}^{\infty}\, {1\over
 {2^c}}\, \sum_{d_1.. d_c}\, \xi^{d_1+ .. + d_a
 + d_{a+1} +.. +d_c}\nonumber \\
 && {{\partial^{c-a}\, \psi^*(\vec x, t)}\over {\partial\, x^{d_{a+1}}\,
 ... \partial\, x^{d_c}}}\, {{\partial^a\, \psi(\vec x, t)}\over {\partial\,
 x^{d_1}\, ... \partial\, x^{d_a}}} =\nonumber \\
 &=& \int d^3\xi\, e^{{i\over {\hbar}}\, \vec p \cdot \vec
 \xi}\, \sum_{c=0}^{\infty}\, {1\over {2^c}}\, \sum_{d_1.. d_c}\, \xi^{d_1+ .. +d_c}\,
 \sum_{a=0}^c\, (-)^a\, \rho_{(c)(d_{a+1} ... d_c)(d_1 .. d_a)}(\vec x,
 t),\nonumber \\
 &&{}\nonumber \\
 &&{}\nonumber \\
 &&\rho_{(c)(d_{a+1} ... d_c)(d_1 .. d_a)}(\vec x, t) =
 {{\partial^{c-a}\, \psi^*(\vec x, t)}\over {\partial\, x^{d_{a+1}}\,
 ... \partial\, x^{d_c}}}\, {{\partial^a\, \psi(\vec x, t)}\over {\partial\,
 x^{d_1}\, ... \partial\, x^{d_a}}},\qquad a > 0,\nonumber \\
 &&\rho_{(c)(d_1 ... d_c)(-)}(\vec x, t) =
 {{\partial^{c}\, \psi^*(\vec x, t)}\over {\partial\, x^{d_1}\,
  ... \partial\, x^{d_c}}}\, \psi(\vec x, t),\qquad a=0,\nonumber \\
 &&\rho_{(c)(-)(d_1 .. d_c)}(\vec x, t) = \psi^*(\vec x, t)\,
 {{\partial^a\, \psi(\vec x, t)}\over {\partial\,
 x^{d_1}\, ... \partial\, x^{d_c}}},\qquad a=c.\nonumber \\
 &&{}
 \label{3.5}
 \eea

By using

\beq
 \int d^3\xi\, e^{{i\over {\hbar}}\, \vec p \cdot \vec
 \xi}\, \xi^{d_1+ .. +d_c}\, = (-i\hbar)^c\, {{\partial^c\, \delta^3(\vec p)}
 \over {\partial\, p^{d_1}\, ... \partial\, p^{d_c}}},
 \label{3.6}
 \eeq

 \noindent and by performing a multipolar expansion of the distribution
 functions $\rho_{(c)(d_{a+1} ... d_c)(d_1 .. d_a)}(\vec x, t) $,

 \bea
 \rho_{(c)(d_{a+1} ... d_c)(d_1 .. d_a)}(\vec x, t) &=&
 \rho_{(c)(d_{a+1} ... d_c)(d_1 .. d_a)o}({\vec x}_{\psi}(t), t)\,
 \delta^3(\vec x - {\vec x}_{\psi}(t)) +\nonumber \\
 &+& \sum_{n=1}^{\infty}\, {{(-)^n}\over {n!}} \sum_{r_1,..,r_n}\,
 \rho_{(c)(d_{a+1} ... d_c)(d_1 .. d_a)n}^{r_1..r_n}({\vec x}_{\psi}(t), t)\,
 {{\partial^n\, \delta^3(\vec x -
 {\vec x}_{\psi}(t))}\over {\partial\, x^{r_1}\, ...\, \partial\,
 x^{r_n}}},\nonumber \\
 &&{}\nonumber \\
 \rho_{(o)o}({\vec x}_{\psi}(t), t) &=&
 \int d^3x\, \rho_{(o)}(\vec x, t) = \rho_{(o)o}(t) = 1,\nonumber \\
 \rho_{(c)(d_{a+1} ... d_c)(d_1 .. d_a)o}({\vec x}_{\psi}(t), t) &=&
 \int d^3x\, {{\partial^{c-a}\, \psi^*(\vec x, t)}\over {\partial\, x^{d_{a+1}}\,
 ... \partial\, x^{d_c}}}\, {{\partial^a\, \psi(\vec x, t)}\over {\partial\,
 x^{d_1}\, ... \partial\, x^{d_a}}} =\nonumber \\
 &=&\rho_{(c)(d_{a+1} ... d_c)(d_1 .. d_a)o}(t),\nonumber \\
  &&(\rho_{(c)(-)(d_1 .. d_c)o}(t))\quad for\, a = c,\nonumber \\
 \rho^{r_1...r_n}_{(c)(d_{a+1} ... d_c)(d_1 .. d_a)n}({\vec x}_{\psi}(t), t)
 &=& \int d^3x\, \Big(x^{r_1} - x_{\psi}^{r_1}(t)\Big)
 ... \Big(x^{r_n} - x_{\psi}^{r_n}(t)\Big)\,
 {{\partial^{c-a}\, \psi^*(\vec x, t)}\over {\partial\, x^{d_{a+1}}\,
 ... \partial\, x^{d_c}}}\, {{\partial^a\, \psi(\vec x, t)}\over {\partial\,
 x^{d_1}\, ... \partial\, x^{d_a}}},\nonumber \\
 &&{}\nonumber \\
 n=1, a=c=0, && \rho^{r_1}_{(o) 1}({\vec x}_{\psi}(t), t) = 0,\,
 for\, EMWF\nonumber \\
 &&{},
 \label{3.7}
 \eea

\noindent we get the following multipolar expansion for the Wigner
function:

\begin{eqnarray*}
 W(\vec x, \vec p, t) &=& W_{(o)}({\vec x}_{\psi}(t), \vec p, t)\,
 \delta^3(\vec x - {\vec x}_{\psi}(t)) +\nonumber \\
 &+& \sum_{n=0}^{\infty}\, {{(-)^n}\over {n!}}\, \sum_{r_1..r_n}\,
  W^{r_1..r_n}_{(o)}({\vec x}_{\psi}(t), \vec p, t)\, {{\partial^n\, \delta^3(\vec x -
 {\vec x}_{\psi}(t))}\over {\partial\, x^{r_1}\, ...\, \partial\,
 x^{r_n}}} =\nonumber \\
 &&{}\nonumber \\
 &{\buildrel {(3.3)}\over =}& \sum_{n=1}^{\infty}\, {{(-)^n}\over
 {n!}}\, \sum_{c=0}^{\infty}\, {{(-i\hbar)^c}\over {2^c}}\,
 \sum_{a=0}^c\, (-)^a\, \sum_{r_1..r_n}\, \sum_{d_1..d_c}\,
 \rho_{(c)(d_{a+1} ... d_c)(d_1 .. d_a)n}^{r_1..r_n}({\vec x}_{\psi}(t), t)\,
 \nonumber \\
 && {{\partial^n\, \delta^3(\vec x -
 {\vec x}_{\psi}(t))}\over {\partial\, x^{r_1}\, ...\, \partial\,
 x^{r_n}}}\,\, {{\partial^c\, \delta^3(\vec p)}
 \over {\partial\, p^{d_1}\, ...  \partial\, p^{d_c}}} =\nonumber \\
 &&{}\nonumber \\
 &=& \sum_{c=0}^{\infty}\, \sum_{d_1..d_c}\, W^{d_1..d_c}_{(o)(c)}({\vec x}_{\psi}(t),
 t)\, {{\partial^c\, \delta^3(\vec p)}
 \over {\partial\, p^{d_1}\, ...  \partial\, p^{d_c}}}\,
 \delta^3(\vec x - {\vec x}_{\psi}(t)) +\nonumber \\
 &+&\sum_{n=1}^{\infty}\, {{(-)^n}\over {n!}}\,\sum_{c=0}^{\infty}\,
 \sum_{d_1..d_c}\, W^{r_1..r_n d_1..d_c}_{(n)(c)}({\vec x}_{\psi}(t),
 t)\, {{\partial^c\, \delta^3(\vec p)}
 \over {\partial\, p^{d_1}\, ...  \partial\, p^{d_c}}}\nonumber \\
 &&{{\partial^n\, \delta^3(\vec x - {\vec x}_{\psi}(t))}\over {\partial\,
 x^{r_1}\, ...\, \partial\, x^{r_n}}},\nonumber \\
 \end{eqnarray*}

\bea
 W_{(o)(o)}({\vec x}_{\psi}(t), t) &=& \rho_{(o)o}({\vec x}_{\psi}(t), t) =
 1,\qquad W^{r_1}_{(1)(o)}({\vec x}_{\psi}(t), t) =
 \rho^{r_1}_{(o) 1}({\vec x}_{\psi}(t), t) = 0,\nonumber \\
 W^{d_1}_{(o)(1)}({\vec x}_{\psi}(t), t) &=& - {{i\hbar}\over 2}\,
 \sum_{a=0}^1 (-)^a \rho_{(1)(d_{a+1})(d_a)o}({\vec x}_{\psi}(t), t)
 =\nonumber \\
 &=& - {{i\hbar}\over 2}\, \int d^3x\,
 \Big(\psi^*\, {{\partial\, \psi}\over {\partial\, x^{d_1}}} - {{\partial\,
 \psi^*}\over {\partial\, x^{d_1}}}\, \psi\Big)(\vec x, t) =
 \rho^{d_1}_{(1)o}(t) = p^{d_1}_{\psi}(t) = m\, {\dot
 x}^{d_1}_{\psi}(t),\nonumber \\
 W^{d_1..d_c}_{(o)(c)}({\vec x}_{\psi}(t), t) &=& (- {{i\hbar}\over 2})^c\,
 \sum_{a=0}^c\, (-)^a\, \rho_{(c)(d_{a+1}..d_c)(d_1..d_a)o}({\vec x}_{\psi}(t),
 t),\nonumber \\
 &&.......
 \label{3.8}
 \eea

The \emph{monopole part} of the Wigner function is

\bea
 W_{(o)}(\vec x, \vec p, t) &=& \sum_{c=0}^{\infty}\,
 \sum_{d_1..d_c}\, W^{d_1..d_c}_{(o)(c)}({\vec x}_{\psi}(t),
 t)\, {{\partial^c\, \delta^3(\vec p)}
 \over {\partial\, p^{d_1}\, ...  \partial\, p^{d_c}}}\,
 \delta^3(\vec x - {\vec x}_{\psi}(t)) =\nonumber \\
 &=& \sum_{c=0}^{\infty}\, {{(-i\hbar)^c}\over {2^c}}\,
 \sum_{a=0}^c\, (-)^a\,  \sum_{d_1..d_c}\,
 \rho_{(c)(d_{a+1} ... d_c)(d_1 .. d_a)n}^{r_1..r_n}({\vec x}_{\psi}(t),
 t)\,\, {{\partial^c\, \delta^3(\vec p)}
 \over {\partial\, p^{d_1}\, ... \partial\, p^{d_c}}}\,
 \delta^3(\vec x - {\vec x}_{\psi}(t)) =\nonumber \\
 &=& \Big(\delta^3(\vec p) + {\vec p}_{\psi}(t) \cdot {{\partial\, \delta^3(\vec p)}
 \over {\partial\, \vec p}} +\nonumber \\
 &+& \sum_{c=2}^{\infty}\, {{(-i\hbar)^c}\over {2^c}}\,
 \sum_{a=0}^c\, (-)^a\,  \sum_{d_1..d_c}\,
 \rho_{(c)(d_{a+1} ... d_c)(d_1 .. d_a)n}^{r_1..r_n}({\vec x}_{\psi}(t),
 t)\, {{\partial^c\, \delta^3(\vec p)}
 \over {\partial\, p^{d_1}\, ... \partial\, p^{d_c}}} \Big)\,
 \delta^3(\vec x - {\vec x}_{\psi}(t)).\nonumber \\
 &&{}
 \label{3.9}
 \eea

This distribution is concentrated in $\vec p = 0$ and $\vec x =
{\vec x}_{\psi}(t)$ so that the problem of not being definite
positive disappears at the monopole level.

\bigskip

The {\it marginals}, i.e. the distribution functions for operators
depending on either the position or the momentum operators only, are

\bea
 W(\vec x, t) &=& \int d^3p\, W(\vec x, \vec p, t) = \rho_{(o)}(\vec x, t)
 =\nonumber \\
 &=& \delta^3(\vec x - {\vec x}_{\psi}(t)) +
 \sum_{n=2}^{\infty}\, {{(-)^n}\over {n!}} \sum_{r_1,..,r_n}\,
 \rho_{(o) n}^{r_1..r_n}({\vec x}_{\psi}(t), t)\, {{\partial^n\, \delta^3(\vec x -
 {\vec x}_{\psi}(t))}\over {\partial\, x^{r_1}\, ...\, \partial\, x^{r_n}}}
  =\nonumber \\
  &=&\delta^3(\vec x - {\vec x}_{\psi}(t)) +
 \sum_{n=2}^{\infty}\, {{(-)^n}\over {n!}} \sum_{r_1,..,r_n}\,
 W_{(o) (n)}^{r_1..r_n}({\vec x}_{\psi}(t), t)\, {{\partial^n\, \delta^3(\vec x -
 {\vec x}_{\psi}(t))}\over {\partial\, x^{r_1}\, ...\, \partial\, x^{r_n}}},\nonumber \\
 &&{}\nonumber \\
 \rho_{(o) n}^{r_1..r_n}({\vec x}_{\psi}(t), t) &=&\rho_{(c)(d_{a+1} ... d_c)(d_1 ..
 d_a)n}^{r_1..r_n}({\vec x}_{\psi}(t), t){|}_{c=a=0},\nonumber \\
 \label{3.10}
 \eea

\bea
 W(\vec p, t) &=&\int d^3x\, W(\vec x, \vec p, t) =\nonumber \\
 &=& {1\over {(2\pi \hbar)^3}}\, \int d^3x_1\, d^3x_2\, e^{{i\over {\hbar}}\, \vec p\cdot
 ({\vec x}_1 - {\vec x}_2)}\, \psi^*({\vec x}_1, t)\, \psi({\vec x}_2,
 t) =\nonumber \\
 &=& \sum_{c=0}^{\infty}\, \sum_{d_1..d_c}\, W^{d_1..d_c}_{(o)(c)}({\vec x}_{\psi}(t),
 t)\, {{\partial^c\, \delta^3(\vec p)}
 \over {\partial\, p^{d_1}\, ...  \partial\, p^{d_c}}}.\nonumber \\
 &&{}
 \label{3.11}
 \eea

\bigskip

For EMWF, we recover $< {\hat {\vec x}}
>_{\psi(t)} = \int {{d^3x\, d^3p}\over {(2\pi\, \hbar)^3}}\, \vec
x\, W(\vec x, \vec p, t) = {\vec x}_{\psi}(t)$ and $< {\hat {\vec
p}} >_{\psi(t)} = \int {{d^3x\, d^3p}\over {(2\pi\, \hbar)^3}}\,
\vec p\, W(\vec x, \vec p, t) = {\vec p}_{\psi}(t) = m\, {\dot {\vec
x}}_{\psi}(t)$.\medskip

It follows that the monopole and dipole of the momentum distribution
${\vec \rho}_{(1)}$ determine the quantities

\bea
 < {\hat x}^i\, {\hat p}^j >_{\psi(t)} &=&  {{i\hbar}\over 2}\,
 \delta^{ij} - x^i_{\psi}(t)\, \rho^j_{(1)o}({\vec x}_{\psi}(t), t)
 + \rho_{(1)1}^{ji}({\vec x}_{\psi}(t), t),\nonumber \\
 < {\hat p}^j\, {\hat x}^i >_{\psi(t)} &=&  - {{i\hbar}\over 2}\,  -
 x^i_{\psi}(t)\, \rho^j_{(1)o}({\vec x}_{\psi}(t), t) +
 \rho_{(1)1}^{ji}({\vec x}_{\psi}(t), t),
 \label{3.12}
 \eea

\noindent in agreement with the commutation relations $[ {\hat
x}^i,\, {\hat p}^j ] =  i\, \hbar\, \delta^{ij}$.\medskip

Moreover, for EMWF, it follows from Eqs.(\ref{3.7})

\bea
 < {\hat x}^i\, {\hat x}^j >_{\psi(t)} &=& x^i_{\psi}(t)\,
 x^j_{\psi}(t) + {1\over 2}\, \rho^{ij}_{(o)2}({\vec x}_{\psi}(t),
 t),\nonumber \\
 < {\hat p}^i\, {\hat p}^j >_{\psi(t)} &=& - \hbar^2\, \rho_{(2)(-)(ij)o}(t).
 \label{3.13}
 \eea

\medskip

Therefore, for the EMWF class, the standard deviations $\triangle\,
x^r = \sqrt{< ({\hat x}^r)^2
>_{\psi(t)} - (<{\hat x}^r>_{\psi(t)})^2}$ of the $\vec x$-distribution
and $\triangle\, p^r = \sqrt{< ({\hat p}^r)^2>_{\psi(t)} - (<{\hat
p}^r>_{\psi(t)})^2}$ of the $\vec p$-distribution, appearing in the
uncertainty relations $\triangle\, x^r\, \triangle\, p^r \geq
\hbar/2$, are determined by the following multipoles :\medskip

$(\triangle\, x^r)^2 = {1\over 2}\,  \rho_{(o) 2}^{rr}({\vec
x}_{\psi}(t), t)$  (the quadrupole of $\rho_{(o)}$);\medskip

$(\triangle\, p^r)^2 = - \hbar^2\, \rho_{(2)(-)(rr)o}(t) -
\Big(\rho^r_{(1)o}(t)\Big)^2 = < ({\hat p}^r)^2>_{\psi} -
\Big(p^r_{\psi}(t)\Big)^2$ (the monopoles of $\rho_{(2)(-)(ij)}$ and
$\rho^r_{(1)o}$ of Eqs.(\ref{3.7}) and (\ref{2.8})).\medskip

As previously said, an energy eigenfunction, i.e. $\Psi_n(\vec x, t)
= e^{{i\over {\hbar}}\, E_n\, t}\, \psi_n(\vec x)$ with $\hat H\,
\psi_n(\vec x) = E_n\, \psi_n(\vec x)$, $\hat H = {{{\hat {\vec
p}}^2}\over {2m}} + V({\hat {\vec x}})$, is a NDWF not of the EMWF
type with a classical trajectory $< {\hat {\vec x}}
>_{\Psi_n(t)} = 0$, if it is a parity eigenstate. For it we get

\bea
 E_n &=& < \hat H >_{\Psi_n(t)} = < \hat H >_{\psi_n} =
 {{< {\hat {\vec p}}^2 >_{\psi_n}}\over {2m}} + < V({\hat {\vec x}})
 >_{\psi_n},\nonumber \\
 &&{}\nonumber \\
 {{< {\hat {\vec p}}^2 >_{\psi_n}}\over {2m}} &=& - {{\hbar^2}\over
 {2m}}\, \sum_r\, \rho_{(2)(-)(rr) o}(t),\nonumber \\
 < V({\hat {\vec x}}) >_{\psi_n} &=& V({\vec x}_{\psi_n}(t))+
 \sum_{m=2}^{\infty}\, {1\over {m!}}\, \sum_{r_1..r_m}\,
 \rho^{r_1..r_m}_{(o)m}({\vec x}_{\psi_n}(t), t)\, {{\partial^m\,
 V({\vec x}_{\psi_n}(t))}\over {\partial\, x^{r_1}_{\psi_n} ..
 \partial\, x^{r_m}_{\psi_n}}}.\nonumber \\
 &&{}
 \label{3.14}
 \eea

For this NDWF the mean kinetic energy  \emph{does not} coincide with
the vanishing classical kinetic energy ${1\over 2}\, m\, {\dot {\vec
x}}^2_{\psi_n}(t) = 0$.

\medskip

Note, finally, that the force ${\vec F}_{\psi_n}(t)$ appearing in
Eq.(\ref{2.9}) is not minus the gradient of the expectation value of
the potential, ${\vec F}_{\psi_n}(t) \not= - {{\partial}\over
{\partial\, {\vec x}_{\psi_n}}}\, < V({\hat {\vec x}}) >_{\psi_n}$,
except when all the higher multipoles  for $n \geq 2$ are
independent of the classical trajectory, i.e.
$\rho^{r_1..r_n}_{(o)n}({\vec x}_{\psi}(t), t) =
\rho^{r_1..r_n}_{(o)n}(t)$.

\vfill\eject

\section{Non-Relativistic Two-Particle Systems}

Let  $\psi({\vec x}_1, {\vec x}_2, t)$ be a two-particle wave
function, ${\vec x}_{1{\psi}}(t)$ and ${\vec x}_{2{\psi}}(t)$ two
classical trajectories. Let us remark that even if $\psi({\vec x}_1,
{\vec x}_2, t)$ lives in the Cartesian product of two Euclidean
spaces, the trajectories ${\vec x}_{1\psi}(t)$ and ${\vec
x}_{2\psi}(t)$ live in the standard Galilei space-time.
\bigskip

The mean values of the position and momentum operators are expressed
in terms of the density matrix $\rho_{(o)}({\vec x}_1, {\vec x}_2,
t) = |\psi({\vec x}_1, {\vec x}_2, t)|^2$ and of two other
probability distributions $\rho^r_{(1)(i)}({\vec x}_1, {\vec x}_2,
t)$

\bea
  < {\hat {\vec x}}_i >_{\psi(t)} &=& \int d^3x_1\, d^3x_2\, {\vec
 x}_i\, \rho_{(o)}({\vec x}_1, {\vec x}_2, t),\nonumber \\
 < {\hat {\vec p}}_i >_{\psi(t)} &=& \int d^3x_1\, d^3x_2\,
 \rho_{(1)(i)}({\vec x}_1, {\vec x}_2, t).
 \label{4.1}
 \eea

and the Ehrenfest theorem implies

\bea
 &&{d\over {dt}}\,  < {\hat {\vec x}}_i >_{\psi(t)}\, = {1\over {m_i}}\,
 < {\hat {\vec p}}_i >_{\psi(t)},\qquad  i=1,2,\nonumber \\
 &&{}\nonumber \\
 &&{d\over {dt}}\, < {\hat {\vec p}}_i >_{\psi(t)}\, = < - {{\partial\,
 V_i({\vec x}_i)}\over {\partial\, {\vec x}_i}} >_{\psi(t)}.
 \label{4.2}
 \eea

\bigskip

The double multipolar expansion of the density matrix is

\begin{eqnarray*}
 \rho_{(o)}({\vec x}_1, {\vec x}_2, t) &=& |\psi({\vec x}_1, {\vec x}_2, t)|^2
 =\nonumber \\
 &=&\rho_{(o) 1o}({\vec x}_{1{\psi}}(t), {\vec x}_2,  t)\, \delta^3({\vec x}_1 - {\vec x}_{1{\psi}}(t))
 +\nonumber \\
 &+& \sum_{n=1}^{\infty}\, {{(-)^n}\over {n!}} \sum_{r_1,..,r_n}\,
 \rho_{(o) 1n}^{r_1..r_n}({\vec x}_{1{\psi}}(t), {\vec x}_2, t)\, {{\partial^n\, \delta^3({\vec x}_1 -
 {\vec x}_{1{\psi}}(t))}\over {\partial\, x_1^{r_1}\, ...\, \partial\,
 x_1^{r_n}}},\nonumber \\
 &&{}\nonumber \\
 n=0,1,..
 &&\rho_{(o) 1n}^{r_1..r_n}({\vec x}_{1{\psi}}(t), {\vec x}_2, t)
 =\nonumber \\
 &=& \rho_{(o) 1o 2o}({\vec x}_{1{\psi}}(t), {\vec x}_{2{\psi}}(t),  t)\,
 \delta^3({\vec x}_2 - {\vec x}_{2{\psi}}(t)) +\nonumber \\
 &+& \sum_{m=1}^{\infty}\, {{(-)^m}\over {m!}} \sum_{s_1,..,s_m}\,
 \rho_{(o) 1n 2m}^{r_1..r_ns_1..s_m}({\vec x}_{1{\psi}}(t),
 {\vec x}_{2{\psi}}(t), t)\, {{\partial^m\, \delta^3({\vec x}_2 -
 {\vec x}_{2{\psi}}(t))}\over {\partial\, x_2^{s_1}\, ...\, \partial\,
 x_2^{s_m}}},\end{eqnarray*}

\bea
 \rho_{(o)}({\vec x}_1, {\vec x}_2, t) &=& |\psi({\vec x}_1, {\vec x}_2, t)|^2
 =\nonumber \\
 &=& \sum_{n,m}^{0..\infty}\, {{(-)^n\, (-)^m}\over {n!\, m!}}\,
 \sum_{r_1..r_ns_1..s_m}\,
 \rho_{(o) 1n 2m}^{r_1..r_ns_1..s_m}({\vec x}_{1{\psi}}(t), {\vec x}_{2{\psi}}(t), t)
 \nonumber \\
 &&{{\partial^n\, \delta^3({\vec x}_1 -
 {\vec x}_{1{\psi}}(t))}\over {\partial\, x_1^{r_1}\, ...\, \partial\,
 x_1^{r_n}}}\, {{\partial^m\, \delta^3({\vec x}_2 -
 {\vec x}_{2{\psi}}(t))}\over {\partial\, x_2^{s_1}\, ...\, \partial\,
 x_2^{s_m}}} =\nonumber \\
 &=&\rho_{(o) 1o 2o}({\vec x}_{1{\psi}}(t), {\vec x}_{2{\psi}}(t),  t)\, \delta^3({\vec x}_1 - {\vec
 x}_{1{\psi}}(t))\, \delta^3({\vec x}_2 - {\vec x}_{2{\psi}}(t)) -\nonumber \\
 &-&\sum_{s_1}\, \rho^{. s_1}_{(o) 1o 21}({\vec x}_{1{\psi}}(t), {\vec x}_{2{\psi}}(t),  t)\,
 \delta^3({\vec x}_1 - {\vec x}_{1{\psi}}(t))\, {{\partial\, \delta^3({\vec x}_2 -
 {\vec x}_{2{\psi}}(t))}\over {\partial\, x_2^{s_1}}} -\nonumber \\
 &-&\sum_{r_1}\, \rho^{r_1 .}_{(o) 11 2o}({\vec x}_{1{\psi}}(t), {\vec x}_{2{\psi}}(t),  t)\,
 {{\partial\, \delta^3({\vec x}_1 - {\vec x}_{1{\psi}}(t))}\over
 {\partial\, x_1^{r_1}}}\, \delta^3({\vec x}_2 -
 {\vec x}_{2{\psi}}(t)) +\nonumber \\
  &+&\sum_{r_1s_1}\, \rho^{r_1s_1}_{(o) 11 21}({\vec x}_{1{\psi}}(t), {\vec x}_{2{\psi}}(t),  t)\,
 {{\partial\, \delta^3({\vec x}_1 - {\vec x}_{1{\psi}}(t))}\over
 {\partial\, x_1^{r_1}}}\, {{\partial\, \delta^3({\vec x}_2 -
 {\vec x}_{2{\psi}}(t))}\over {\partial\, x_2^{s_1}}} + ....\nonumber \\
 &&{}
  \label{4.3}
 \eea

\bigskip

The condition $<{\hat {\vec x}}_1>_{\psi(t)} = {\vec
x}_{1{\psi}}(t)$ requires the vanishing of the dipole $\rho^{r_1
.}_{(o) 11 20}({\vec x}_{1{\psi}}(t), {\vec x}_{2{\psi}}(t), t) =
0$, while $< {\hat {\vec x}}_2 >_{\psi(t)} = {\vec x}_{2{\psi}}(t)$
requires the vanishing of the dipole $\rho^{. s_1}_{(o) 10 21}({\vec
x}_{1{\psi}}(t), {\vec x}_{2{\psi}}(t),  t) = 0$. If the third
dipole vanishes too, i.e. $\rho^{r_1s_1}_{(o) 11 21}({\vec
x}_{1{\psi}}(t), {\vec x}_{2{\psi}}(t), t) = 0$, it follows $<{\hat
x}^{r}_1\, {\hat x}^{s}_2>_{\psi(t)} = x^r_{1{\psi}}(t)\,
x^s_{2{\psi}}(t)$.\medskip

For EMWF, for which the Ehrenfest theorem holds, by making a
multipolar expansion of $\rho^r_{(1)(i)}({\vec x}_1, {\vec x}_2,t) =
\Big(\psi^*\, i\hbar\, {{\partial\, \psi}\over {\partial\,
x_i^r}}\Big)({\vec x}_1, {\vec x}_2,t)$, from the monopole terms
$\rho^r_{(1)(i)o}(t)$, due to the first of Eqs.(\ref{4.2}), we get
\medskip

\beq
 < {\hat {\vec p}}_i >_{\psi(t)} = m_i\, {\dot {\vec x}}_{i
 \psi}(t).
 \label{4.4}
 \eeq

\medskip

With potentials $V_i({\vec x}_1, {\vec x}_2) = V_{12}({\vec x}_1 -
{\vec x}_2) + V_{ext}({\vec x}_i)$, if all dipoles vanish, Eqs
(\ref{4.2}) and (\ref{4.3}) imply the following coupled
deterministic equations of motion for the two trajectories

\bea
 m_i\, {\ddot {\vec x}}_{i \psi}(t) &=& - {{\partial\, V_i({\vec x}_{1
 \psi}(t), {\vec x}_{2 \psi}(t))}\over {\partial\, {\vec x}_{i
 \psi}}} -\nonumber \\
 &-& \sum_{n,m}^{2..\infty}\, {{(-)^n\, (-)^m}\over {n!\,
 m!}}\, \sum_{r_1..r_ns_1..s_m}\, \rho_{(o) 1n
 2m}^{r_1..r_ns_1..s_m}({\vec x}_{1{\psi}}(t), {\vec x}_{2{\psi}}(t),
 t)\nonumber \\
 && {{\partial^{n+m}\, V_i({\vec x}_{1 \psi}(t), {\vec x}_{2
 \psi}(t))}\over {\partial\, {\vec x}_{i \psi}\, \partial\,
 x^{r_1}_{1 \psi}... \partial\, x^{r_n}_{1 \psi}\, \partial\,
 x^{s_1}_{2 \psi}.. \partial\, x^{s_m}_{2 \psi}}}.
 \label{4.5}
 \eea

\medskip

On the other hand, if only one dipole vanishes, for instance
$\rho^{r_1 .}_{(o) 11 20}({\vec x}_{1{\psi}}(t), {\vec
x}_{2{\psi}}(t), t) = 0$, we have only $< {\hat {\vec x}}_1
>_{\psi(t)} = {\vec x}_{1{\psi}}(t)$ and we can write a
deterministic equation of motion for particle 1 but not for particle
2.

\bigskip

Therefore, by means of a double multipolar expansion, it is possible
to find solutions of the Schroedinger equation with two {\it
effective} classical particles, solutions with only one {\it
effective }classical particle and, in general, solutions without
{\it effective} classical particles.

\medskip

Let us stress that for a two-particle quantum system there is a
further ambiguity in the position basis of Hilbert space. Actually,
according to the standard notion of separability, the Hilbert space
is described as the tensor product $H = H_1 \otimes H_2$, so that a
wave function is separable if $\psi({\vec x}_1,{\vec x}_2) =
\psi_1({\vec x}_1)\, \psi_2({\vec x}_2)$ and entangled if
$\psi({\vec x}_1,{\vec x}_2) \not= \psi_1({\vec x}_1)\, \psi_2({\vec
x}_2)$.\medskip

However, in the case of an isolated system of two interacting
particles, the only way of solving the Schroedinger equation is to
transform to center-of-mass  and relative variables. Then, by means
of a unitary transformation, the Hilbert space is represented as $H
= H_{c.m} \otimes H_{rel}$. Now, as always happens in scattering
theory, a wave function is separable if $\Psi(\vec x, \vec r) =
\Psi_{c.m}(\vec x)\, \Psi_{rel}(\vec r)$ and entangled (note that
this is a totally different notion of entanglement compared to one
mentioned above) if $\Psi(\vec x, \vec r) \not= \Psi_{c.m}(\vec x)\,
\Psi_{rel}(\vec r)$. In scattering theory, however, one never
considers wave packets with superpositions of the center of mass.
The total momentum is a constant of motion and most likely there is
some kind of super-selection rule forbidding these
superpositions.\medskip

Now the center of mass is a free quantum particle, while the
relative quantum motion can be either {\it effectively} 'classical'
(zero dipole) or not. The free center of mass is not associated with
an {\it effective} 'classical' particle, because it is described by
a plane wave (non-zero dipole) having the momentum of the isolated
two-particle system.

\bigskip

N particle systems can be treated in the same way.

\vfill\eject

\section{Relativistic Quantum Mechanics in the Rest Frame Instant Form
of Dynamics}

The extension of our approach to special relativity needs a
consistent formulation of RQT in inertial frames of Minkowski
space-time, a corresponding Ehrenfest theorem, and a well-defined
non-relativistic limit to QT in the inertial frames of Galilei
space-time.\medskip

In non-relativistic physics, a classical theory of particles in
Galilean space-time (where both time and 3-space are absolute
notions) is usually worked out with the position of particles as
basic preferred observables.
\medskip

Newtonian classical mechanics is 'local' in a {\it strict sense} if
external potentials acting on the particles are admitted but
inter-particle (instantaneous) potentials $V({\vec x}_i - {\vec
x}_j)$ are not ('non-local' case). Both in the 'local' and
'non-local' cases there is no problem with the definition of the
Galilean \emph{center of mass}. The Galilei generators (constants of
motion for isolated systems) are 'local' and only the Galilei energy
possibly depends on 'non-local' potentials $V({\vec x}_i - {\vec
x}_j)$. The Galilei boosts do not depend on the potentials and can
be used to define the Newtonian center of mass (unique canonical and
covariant collective position variable), which is assumed to be a
'measurable' quantity. The above kinds of 'non-locality' are
harmless in the non-relativistic theory. One can safely define a
sub-system of an isolated system at least in absence of non-local
potentials.
\medskip

This viewpoint becomes problematic in special relativity and we will
see that although relativistic classical mechanics of isolated
systems in Minkowski space-time hatches the same two sources of
'non-locality' as the non-relativistic case, in a relativistic
framework these give way to significant consequences.\medskip

\subsection{Problems of the Relativistic Framework}

The lack of a well-defined quantum-relativistic framework,
independent of quantum field theory yet compatible with it (e.g.
such to allow a description of charged particles interacting with
the electro-magnetic field) is at the origin of many problems for
the formulation of relativistic extensions of various existing
proposals for the solution of the quantum measurement problem.

\medskip

The basic problems affecting this attempt, essentially residing in
the Lorentz signature of Minkowski space-time, gave rise in the last
century to a vast literature, starting from Fokker (see
Ref.\cite{27,28,29} and the bibliography therein). They are:
\medskip

A) The lack of an absolute notion of 3-space and the ensuing
necessity to choose a convention for the synchronization of distant
clocks.\medskip

B) The structure of the Poincar\'e group. While in the realizations
of the (mass-)extended Galilei algebra the energy generator only
depends on the mutual particle interaction (sum of the kinetic and
the potential energy), in the (Dirac) {\it instant form} of
relativistic dynamics both the energy and the Lorentz boost
generators are interaction dependent. Since the actual structure of
the Poincar\'e generators follows from the energy-momentum tensor of
an isolated system, such generators are consequently 'non-local'
quantities in the sense of requiring information about the whole
3-space $\Sigma_{\tau}$ identified by the clock synchronization
convention.

C) The lack of a collective variable for an isolated system having
all the properties of the Newton center of mass (see
Refs.\cite{15,16,17,29}). As is well known, the only intrinsic (i.e.
containing information about the isolated system alone) relativistic
collective variables (replacing the Newtonian center of mass and
tending to it in the non-relativistic limit) are the canonical
non-covariant Newton-Wigner center of mass (or center of spin), the
non-canonical covariant Fokker-Pryce center of inertia and the
non-canonical non-covariant M$\o$ller center of energy. All these
can be built in terms of the Poincare' generators alone. They are
all {\it 'non-local', non measurable} quantities due to the
'non-locality' of the generators.

\medskip

D) In isolated systems 'non-local' potentials must depend on {\it
space-like relative variables} only. This is a consequence of {\it
the elimination of relative times in the theory of relativistic
bound states to avoid time-like levels not present in spectroscopic
data}. Consequently, even in the free case, sub-systems of the
isolated system are kinematically interconnected in the physical
Hilbert space ('spatial non-separability'; 'weak relationism'; no
intrinsic definition of subsystems). This kinematical 'non-locality'
allows the presence of 'non-local' potentials of the Coulomb or
Darwin type, e.g. in the treatment of systems of charged particles
plus the electro-magnetic field (after the gauge fixation of the
un-physical gauge variables of the electro-magnetic field). Such
'non-locality' is allowed by special relativity so that, given the
Cauchy data, no problem arises concerning time evolution.
\bigskip

Usually, QT is considered 'local' in the sense of the no-signalling
theorem and 'non-local' for the EPR correlations among sub-systems.
\footnote{This issue is still debated, as can be seen in the last
Section of Ref.\cite{24}}.

\medskip

Let us stress that the 'kinematical non-locality' and 'spatial
non-separability' connected with Lorentz signature imply that the
relativistic description is intrinsically {\it 'non-local' in the
previous sense} {\it before} taking into account the {\it specific
sense} of the quantum EPR 'non-locality'. We shall refer to these
two kinds of 'non-locality' appearing in relativistic quantum
formulations as 'kinematical' and 'causal', respectively.
\medskip

Let us note, incidentally, that the 'kinematical non-locality' must
be taken into account at the non-relativistic level if the $1/c$
coupling to the electro-magnetic field must be taken into account,
as in atomic and molecular physics.
\medskip

\subsection{Relativistic Quantum Mechanics}

In Ref.\cite{16} a consistent formulation of RQT is expounded which
takes into account all the known problems implied by the Lorentz
signature. This formulation is obtained as the restriction to
inertial frames of the {\it parametrized Minkowski theories} of
Ref.\cite{27}, \cite{17}, describing isolated systems in arbitrary
admissible non-inertial frames (see Ref.\cite{28}). This inertial
formulation, {\it the rest-frame instant form of dynamics}, is
originally developed in the rest-frame of the isolated system (and
then re-expressed in arbitrary inertial frames) after the decoupling
of the external, non-covariant, non-local, non-measurable canonical
center of mass. The clock synchronization convention corresponds to
a 3+1 splitting of space-time centered on the Fokker-Planck center
of inertia of the system (as an observer) and exploits radar
4-coordinates: a proper time $\tau$ and 3-coordinates having the
observer as origin. In the resulting Euclidean so-called Wigner
3-spaces of the rest frame, N positive-energy particles are
described by N Wigner spin-1 position 3-vectors ${\vec
\eta}_i(\tau)$, i=1,..,N, and by 3-momenta ${\vec \kappa}_i(\tau)$.
$\sum_i\, {\vec \kappa}_i \approx 0$ is the rest-frame condition and
the 3-coordinates of the internal 3-c.m. are eliminated by the
vanishing of the internal (interaction-dependent) Lorentz boosts (in
the free case we have $\sum_i\, {\vec \eta}_i\, \sqrt{m_i^2c^2 +
{\vec \kappa}_i^2} \approx 0$). Therefore, here too, only relative
variables survive.
\medskip

The world-lines $x^{\mu}_i(\tau)$ of the particles are derived
(interaction-dependent) quantities and in general do not satisfy
vanishing Poisson brackets: note then that at the classical level a
kind of {\it non-commutative structure} already emerges due to the
Lorentz signature of space-time. The world-lines are built in terms
of the Fokker-Pryce center-of-inertia 4-vector $Y^{\mu}(\tau)$, the
conserved 4-momentum $P^{\mu}$ of the isolated system and the
3-positions ${\vec \eta}_i(\tau)$, as shown in Ref.\cite{16}
$x^{\mu}_i(\tau) = Y^{\mu}(\tau) + \epsilon^{\mu}_r(\vec h)\,
\eta^r_i(\tau)$ \footnote{Where $\vec h = \vec P/\sqrt{\sgn\, P^2}$
and $\epsilon^{\mu}_r(\vec h)$ are the spatial columns of the
standard Wigner boost bringing the time-like 4-vector $P^{\mu}$ at
rest.}. The standard 4-momenta $p^{\mu}_i(\tau)$, satisfying $p^2_i
= m_i^2\, c^2$ in the free case, are derived quantities too.

\bigskip

As shown in Ref.\cite{16}, taking into account the elimination of
the relative times needed for a consistent treatment of relativistic
bound states \cite{28,30,31} requires the description of RQT in a
Hilbert space of the type $H = H_{c.m} \otimes H_{rel}$ where
$H_{rel}$ is the Hilbert space of the relative variables. $H_{c.m}$
is the Hilbert space of the decoupled non-covariant relativistic
center of mass, which must be described in terms of frozen Jacobi
data to avoid Hegerfeldt theorem (instantaneous spreading of c.m.
wave packets). These Jacobi data are described by a self-adjoint
velocity operator ${\hat {\vec h}}$ and by an operator ${\hat {\vec
z}}$, whose classical expression is the product of the rest mass of
the isolated system for the initial value of the classical analogue
of the Newton-Wigner position operator. As explained in Appendix A,
this operator and therefore even ${\hat {\vec z}}$, cannot be taken
as self-adjoint. This implies bad localization of this non
measurable quantity. The operator ${\hat Y}^{\mu}$, corresponding to
the classical Fokker-Pryce center of inertia, is defined in terms of
the quantum Jacobi data and of the quantum internal mass and rest
spin after the choice of a suitable operator ordering.

\medskip

Because of the elimination of relative times, this Hilbert space is
{\it not} unitarily equivalent to the standard separable Hilbert
space $H_1 \otimes H_2 \otimes ...$ built in terms of the single
particle Hilbert spaces $H_i$, like in the non-relativistic case. It
is seen that, unlike the standard non-relativistic situation, the
Lorentz signature implies not only {\it kinematical non-locality}
but also {\it non separability}.

\medskip

The non-relativistic limit of the RQT given in Ref.\cite{16}
reproduces the standard QT in the center-of-mass  and relative
variable configuration representation after a unitary transformation
to the Hamilton-Jacobi description of the Newton center of mass
(frozen center of mass limit of the relativistic Jacobi data). Since
atomic and molecular physics require the presence of the
electromagnetic field at the order $1/c$ (at this order neither the
Galilei nor the Poincar\'e group are implemented), in such
experimental situations, even non-relativistic QT, being the $c
\rightarrow \infty$ limit of RQT, should be presented in this
way\footnote{For a review of the results obtained so far along these
lines, see Ref.\cite{15}, where the treatment of the gravitational
field (at the classical level) is also given.}.

\bigskip

While in QT one can consider one particle with external interaction
(with non-conserved Galilei generators), in RQT if we expect to
maintain the Poincar\'e generators and the kind of interactions
compatible with Poincar\'e algebra \cite{28,30,31} under control, a
one particle system must be treated as an open subsystem of an
isolated N-particle system. \medskip

Here, we will therefore consider an isolated system of two free
particles as the simplest non trivial model for studying the
transition from the quantum to the classical relativistic world.
Interactions can be introduced along the lines of
Refs.\cite{28,30,31}. After separating the external free decoupled
center of mass and having satisfied the rest-frame conditions, we
get a Schroedinger equation for a {\it Lorentz scalar} wave function
$\psi(\vec \rho, \tau)$ on the Wigner 3-space, depending only on the
relative variable $\vec \rho = {\vec \eta}_1 - {\vec \eta}_2$ in the
configuration representation\footnote{In the interacting case there
are two types of interacting classes with a complete control of the
Poincar\'e algebra \cite{28,30,31}: A) $\hat H = \sum_{i=1}^2\,
\sqrt{m_i^2\, c^2 + V({\hat {\vec \rho}}^2) + {\hat {\vec \pi}}^2}$
with arbitrary potential $V$; B) $\hat H = \sum_{i=1}^2\,
\sqrt{m_i^2\, c^2  + {\hat {\vec \pi}}^2} + V({\hat {\vec
\rho}}^2)$, but only for special potentials like Coulomb plus
Darwin.}, namely (note that scalar products of Wigner 3-vectors are
Lorentz scalars):
\medskip

\beq
 i \hbar {{\partial}\over {\partial\, \tau}}\, \psi(\vec \rho, \tau)
 = \Big(\sqrt{m_1^2\, c^2 + {\hat {\vec \pi}}^2 } + \sqrt{m_2^2\, c^2
 + {\hat {\vec \pi}}^2 }\Big)\, \psi(\vec \rho, \tau) = \hat H\,
 \psi(\vec \rho, \tau),
 \label{5.1}
 \eeq

\noindent where ${\hat {\vec \pi}} = i \hbar \, {{\partial}\over
{\partial\, \vec \rho}}$ is the relative momentum operator conjugate
to the relative position operator ${\hat {\vec \rho}}$.

\medskip

Modulo ordering problems, the operators ${\hat {\vec \eta}}_i$, i =
1,2, associated with the localization of the two particles inside
the Wigner 3-spaces, can be expressed as well defined functions of
${\hat {\vec \rho}}$ and ${\hat {\vec \pi}}$ starting from the
classical expressions given in Ref.\cite{28}.

\medskip

A first technical problem with Eq.(\ref{5.1}) is the necessity of
exploiting pseudo-differential operators \cite{32} in order to give
a meaning to quantities like $\sqrt{m_i^2\, c^2 + {\hat {\vec
\pi}}^2 }$.\medskip

With this technique we can study the space of solutions of the
relativistic Schroedinger equation. We can associate a density
function $\rho_{(o)}(\vec \rho, \tau) = |\psi(\vec \rho, \tau)|^2$
and the general Wigner distribution function to each normalizable
solution, like in the non-relativistic case. We can then define
their multipolar expansions around a classical trajectory ${\vec
\rho}_{\psi}(\tau)$, which at any instant of proper time $\tau$
defines a relative position 3-vector in the Wigner 3-space
$\Sigma_{\tau}$.

\medskip

A problem with such pseudo-differential operators in the quantum
Hamiltonian is the necessity of characterizing the function space
needed as Hilbert space in such a way that a {\it relativistic
version of the Ehrenfest theorem} holds for the expectation values
of the relative position and relative momentum operators, as a
consequence of the Heisenberg equations ${{d {\hat {\vec \rho}}
}\over {d\tau}} = i \hbar\, [ {\hat {\vec \rho}}, \hat H ]$, ${{d
{\hat {\vec \pi}} }\over {d\tau}} = i \hbar\, [ {\hat {\vec \pi}},
\hat H ]$. In the free case this would imply
\medskip

\bea
 &&{d\over {d\tau}}\, < {\hat {\vec \rho}} >_{\psi(\tau)} = < {\hat
 {\vec v}}\, {\buildrel {def}\over =}\, {\hat {\vec \pi}}\,
 \Big({1\over {\sqrt{m_1^2\, c^2 + {\hat {\vec \pi}}^2 }}} +  {1\over
 {\sqrt{m_2^2\, c^2 + {\hat {\vec \pi}}^2 }}} \Big)>_{\psi(\tau)},
 \nonumber \\
 &&{}\nonumber \\
 &&{d\over {d\tau}}\, < {\hat {\vec \pi}} >_{\psi(\tau)} = 0,
 \label{5.2}
 \eea

\noindent where ${\hat {\vec v}}$ is the 3-velocity operator. Thus,
in this case, we can invert the relation between ${\hat {\vec v}}$
and ${\hat {\vec \pi}}$ to get  ${\hat {\vec \pi}} = {{m \, {\hat
{\vec v}}}\over {\sqrt{4 - {\hat {\vec v}}}}}$ in the equal mass
case, recovering the relativistic standard connection between
3-momentum and 3-velocity.

\medskip

If the dipole of the density matrix vanishes, we get $< {\hat {\vec
\rho}} >_{\psi(\tau)} = {\vec \rho}_{\psi}(\tau)$. Then for EMWFs
from the first half of Ehrenfest theorem we obtain

\bea
 {\dot {\vec \rho}}_{\psi}(\tau) &=& < {\hat {\vec v}}
 >_{\psi(\tau)},\nonumber \\
 &&{}\nonumber \\
 \Rightarrow &&
 < {\hat {\vec \pi}} >_{\psi(\tau)} = {{m\, {\dot {\vec
 \rho}}_{\psi}(\tau)}\over {4 - {\dot {\vec \rho}}^2_{\psi}(\tau)}},
 \qquad (equal\, mass\, case).
 \label{5.3}
 \eea

\noindent while the second part of Ehrenfest theorem provides the
deterministic second order relativistic equation  ${\ddot {\vec
\rho}}_{\psi}(\tau) = 0$.

\bigskip

In presence of interactions, the calculations become very complex.
We see, however, that the emergence of 'classical' results for EMWF
follows the same pattern of the non-relativistic case, in the
center-of-mass - relative configuration representation for the RQT
of Ref.\cite{16}.
\medskip

In conclusion, for a 2-particle system  only the relative variable
in the Wigner 3-spaces can sustain a classical behavior in the sense
that, if the dipole of the density matrix vanishes and the wave
function is of the EMWF type, the system can described by a
'classical' relative variable.\medskip

On the other hand, the decoupled external ('non-local',
'non-measurable') center of mass is described by a plane wave
corresponding to the conserved momentum of the isolated system and,
like in the non-relativistic scattering cases, it does not admit a
classical visualization.
\medskip

The reconstruction of effective classical world-lines requires the
study of the expectation value of the operators ${\hat x}^{\mu}_i =
{\hat Y}^{\mu} + \epsilon^{\mu}_r({\hat {\vec h}})\, {\hat
\eta}^r_i$ among wave functions whose part describing the relative
motion is an EMWF. This is a difficult task which we intend to deal
with in a future paper, trying to take into account the problems of
localization described in Appendix A. In any case, the experimental
determination of the world-line will lead to the localization of
{\it effective} relativistic massive particles in a world-tube with
a radius of the order of nanometers, i.e. much larger than the
Compton wavelength of the isolated system.

\bigskip

The extension to N-particle systems is straightforward.

\vfill\eject

\section{On the Possible Relevance of Classical 'Effective' Trajectories in
Actual Experiments Involving Massive Particles}

As previously stated, we believe that the proposed notion of
classical 'emergence'- specifically the {\it effective} monopole
massive particles - justifies the actual efficacy of the
quasi-classical Newtonian intuition implicitly used by atomic and
molecular physicists in devising experiments and describing the
properties of complex systems. This is true, in particular, for the
experimental description of the preparation, the intermediate steps
and the final outcome of experiments in particle, nuclear, atomic
and molecular physics, where the 'particle' aspect dominates.
Actually in the {\it preparation} of all the existing experiments
there are classical macroscopic apparatuses creating {\it a beam of
quantum particles with a mean classical trajectory and a mean
momentum}, described in terms of Newtonian or relativistic beams of
{\it effective} quasi-classical particles. This is true from
particle accelerators to atoms localized in a cavity. We argue that
the wave functions describing the input beams should have small
dipoles approximating EMWFs. It seems that no conceivable experiment
can avoid, at a certain level of approximation, this quasi-classical
description. Note that the preparation stage should often be
considered independently of the type of measurement to be made in
the subsequent stages of the experiment.\medskip

Let us remark that in all the experiments in which the relevant
system is a 'quantum massive particle', notwithstanding the
criticism of the POVM approach \cite{bb}, we meet the so-called
problem of the 'wave-particle' dualism: a single electron in QT is,
strictly speaking, neither a particle nor a wave, but sometimes its
'particle aspects' dominate while in other cases the 'wave aspects'
do so. Both the 'particle aspects' and the 'wave aspects' are
summarized by the wave function of the electron and become dominant
in different experiments.

\medskip

All this is of course an idealization. As already observed, the
preparation of both scattering experiments (particles at CERN,
atoms, molecules) and the double slit experiment require collimated
beams of effective particles and the effective mean momentum (and
mean energy) of the beam is defined with time-of-travel methods (in
Refs. \cite{bb} it is said that this quantity does not derive from
the momentum operator of a quantum particle, which instead is needed
for the indetermination relations). It is important to stress that
only when the mean energy of a collimated beam is below a certain
threshold we can speak of an 'effective one particle' in the beam.

\medskip

In what follows we discuss atom interferometers and the double slit
experiment as two relevant instantiations of the 'particle' and
'wave' aspects.

\subsection{ Effective Particles in the Arms of Atom Interferometers
from Feynman Path Integral}

Atom interferometry is a rapidly developing area of research in
which interference effects ($\psi = \psi_1 + \psi_2$ with $\psi_i$
describing the atom beam in the arm $i$) compatibly coexist with a
quasi-classical description of the beams in the two arms, in terms
of \emph{effective} atoms, based on Feynman path integral approach.
As shown in Refs.\cite{33} this is possible because in each arm the
dominating classical trajectory generates a classical action
$S_{cl}({\vec x}_f, t_f; {\vec x}_i, t_i) >> \hbar$. This classical
action is the main tool for determining the phases inside the atom
interferometer and for generating the classical trajectories that,
in presence of Newtonian gravity, allow for instance to perform
measures of $g$ and test the universality of free fall.\footnote{See
the discussion on the possibility of testing the gravitational
redshift with quantum atoms in Refs.\cite{34}.}.

\medskip

Our results are compatible with this picture, suggesting that the
wave functions in the two arms have associated density matrices
$|\psi_i|^2$ with {\it small dipole} and almost EMWF wave functions:
this is a condition on the {\it modulus} of the path integral and
not on its {\it phase}. The non-Newtonian forces we get for
non-quadratic potentials should be connected with higher order terms
beyond the classical action.

\subsection{Wave and Particle Aspects in the Double Slit Experiment}

In the double slit experiment with massive particles one
traditionally emphasizes their wave aspect mainly due to the
importance of the interference effects involved. However, particle
aspects must be considered too for a satisfactory account of the
whole phenomenon. In order to clarify this point we shall make
recourse to some mathematical aspects of Bohmian mechanics,
disregarding its ontological interpretation. Note that our approach
helps to clarify aspects of the {\it problem of the classical
regime} without any implication concerning interpretational problems
of QT.

\medskip

The starting point of the 'pilot wave' description of QT \cite{8,9}
is a rewriting of the Schroedinger equation in a form reminiscent of
the eikonal approximation of classical electro-magnetic waves, where
a wave is represented as a wave front and a congruence of rays (the
light rays, trajectories of unspecified massless photons) orthogonal
to it\footnote{As well-known, according to Bohmian ontology, such
effective rays are interpreted as Bohmian trajectories of added
'real' particles.}. The wave function of the quantum massive
particle is written in the form

\beq
 \psi(\vec x, t) = \sqrt{\rho_{(o)}(\vec x, t)}\, e^{{i\over {\hbar}}\, S(\vec x, t)},
 \qquad \rho_{(o)} = |\psi|^2,
 \label{6.1}
 \eeq

\noindent where $\rho_{(o)}$ is the density matrix and $S$ an action
variable.\medskip

Then the Schroedinger equation with Hamiltonian ${{{\vec p}^2}\over
{2m}}\, + V$ is rewritten as the following pair of coupled
equations:\medskip

1) A guidance equation that is the continuity equation (\ref{2.1})
$\partial_t\, \rho_{(o)}(\vec x, t) = \vec \partial \cdot {\vec
\rho}_{(1)}(\vec x, t)$. The Bohm velocity field, i.e. the field of
the tangent vectors to the effective rays, is

\beq
 {\vec v}_B(\vec x, t) = {{{\vec \rho}_{(1)}(\vec x, t)}\over {\rho_{(o)}(\vec x, t)}}
 = {{\hbar}\over m}\, Im\, {{\vec \partial\, \psi}\over {\psi}}(\vec
 x, t) = {1\over m}\, \vec \partial\, S(\vec x, t).
 \label{6.2}
 \eeq

\noindent The Bohm velocity field is the velocity field of a fluid
of hydrodynamic type whose fluid elements follow the Bohmian
trajectories. In the double slit experiment, the initial points can
be any of the points inside the two slits.

\medskip

2) An equation of the Hamilton-Jacobi type for the action $S(\vec x,
t)$

\beq
 - \partial_t\, S(\vec x, t) = {1\over {2m}}\, \Big({\vec \partial}^2\,
 S(\vec x, t)\Big)^2 + V(\vec x, t) + Q(\vec x, t),
 \label{6.3}
 \eeq

\noindent where $Q(\vec x, t)$ is the so-called \emph{quantum
potential} and has the following expression

\beq
 Q(\vec x, t) = - {{\hbar^2}\over {2m}}\, {{ {\vec \partial}^2\,
 \sqrt{\rho_{(o)}(\vec x, t)} }\over {\sqrt{\rho_{(o)}(\vec x,
 t)}}}.
 \label{6.4}
 \eeq

\noindent

Note that Eq.(\ref{6.3}) resembles the Hamilton-Jacobi equation for
a classical material particle.\medskip

Eqs.(\ref{6.2}) and (\ref{6.3}) imply a quasi-Newtonian equation of
motion for the fluid similar to the classical Euler equation

\bea
 m\, {d\over {dt}}\, {\vec v}_B(\vec x, t) &=& m\, \Big(\partial_t +
 {\vec v}_B(\vec x, t) \cdot \vec \partial\Big)\,
 {\vec v}_B(\vec x, t) =\nonumber \\
 &=& - \vec \partial\, \Big(V(\vec x, t) + Q(\vec x, t)\Big).
 \label{6.5}
 \eea

 \bigskip

Independently of Bohm's ontological interpretation, if ${\vec
x}_B(t)$ is formally a Bohmian trajectory, we can use the multipolar
expansions (\ref{2.4}) and (\ref{2.5}) around it in Eq.(\ref{6.2})
to get the following expression

\bea
   {\vec \rho}_{(1)}(\vec x, t) &=& {\vec \rho}_{(1)o}({\vec x}_{\psi}(t),
 t)\, \delta^3(\vec x - {\vec x}_{\psi}(t))
 + \sum_{n=1}^{\infty}\, {{(-)^n}\over {n!}} \sum_{r_1,..,r_n}\,
 {\vec \rho}{}_{(1)n}^{r_1..r_n}({\vec x}_{\psi}(t), t)\, {{\partial^n\, \delta^3(\vec x -
 {\vec x}_{\psi}(t))}\over {\partial\, x^{r_1}\, ...\, \partial\, x^{r_n}}} =\nonumber \\
 &&{}\nonumber \\
 &=& \vec \partial\, S(\vec x, t)\, \rho_{(o)}(\vec x, t) =\nonumber \\
 &=& \vec \partial\, S(\vec x, t)\, \Big(
 \rho_{(o) o}({\vec x}_{\psi}(t), t)\, \delta^3(\vec x - {\vec x}_{\psi}(t))
 +\nonumber \\
 &+& \sum_{n=1}^{\infty}\, {{(-)^n}\over {n!}} \sum_{r_1,..,r_n}\,
 \rho_{(o) n}^{r_1..r_n}({\vec x}_{\psi}(t), t)\, {{\partial^n\, \delta^3(\vec x -
 {\vec x}_{\psi}(t))}\over {\partial\, x^{r_1}\, ...\, \partial\, x^{r_n}}}\Big),\nonumber \\
 &&{}
 \label{6.6}
 \eea

\noindent whose monopole term is (Eq.(\ref{2.6}) is used)

\bea
 {\vec \rho}_{(1)o}({\vec x}_B(t), t)
 &=& < {\hat {\vec p}}_{\psi(t)} > =
 m\, {\vec v}_B({\vec x}_B(t), t) +\nonumber \\
  &+& m\, \sum_{n = 1}^{\infty}\, {{(-)^n}\over {n!}} \sum_{r_1,..,r_n}\,
 \rho_{(o) n}^{r_1..r_n}({\vec x}_B(t), t)\,
 {{\partial^n\, \vec \partial\, S({\vec x}_B(t), t)}\over
 {\partial\, x_B^{r_1}\, ...\, \partial\,
 x_B^{r_n}}}.
 \label{6.7}
 \eea

No statement can be made, at this stage, on the vanishing of the
dipole term ${\vec \rho}_{(o)1}({\vec x}_B(t), t)$ of the density
matrix.
\medskip

According to Bohm's ontology, the effective rays are real particles
hitting on the detector so that - {\it in our interpretation} - near
that point of the detector, the Bohmian trajectory should be
reinterpreted as an \emph{effective} monopole particle with ${\vec
\rho}_{(o)1}({\vec x}_B(t), t) = 0$, i.e. having vanishing dipole
and being an EMWF. Now, when is this picture valid?

\bigskip

Consider the double slit experiment. The Bohmian trajectories
emanating from the two slits correspond to a fluid whose fluid
trajectories rarefy in the spatial regions corresponding to the dark
fringes on the screen, where the wave function is nearly zero or of
order $O(\hbar)$ (if it is represented as a Feynman path integral
this means that there is destructive interference between the
Feynman paths).\medskip

On the other hand, in the regions corresponding to the bright
fringes of the screen there is a concentration of fluid
trajectories. In each point of this region the density matrix is
near a maximum (where we have $\rho_{(o)}(\vec x, t) \approx const.$
with all spatial derivatives small), around which the quantum
potential $Q({\vec x}_B(t))$ of Eq.(\ref{6.4}) is very small (of
order $O(\hbar^2)$), so that Eq.(\ref{6.3}) is approximately the
Hamilton equation of a classical particle. Then $S({\vec x}_B(t), t)
\approx S_{clas}({\vec x}_{clas}(t), t)$, where $S_{clas}$ is a
classical action evaluated along a classical trajectory ${\vec
x}_{clas}(t) \approx {\vec x}_B(t)$. By exploiting the stationary
phase approximation, one finds that in such points of the screen,
the Feynman path integral representation of the double slit wave
function \footnote{See for instance Ref.\cite{35} for the double
slit experiment.} is dominated by the path ${\vec x}_{clas}(t)$ and
has the form $\psi({\vec x}_{clas}(t), t) \approx
\sqrt{\rho_{(o)}({\vec x}_{clas}(t), t)}\, e^{{i\over {\hbar}}\,
S_{clas}({\vec x}_{clas}(t), t)}\, \approx const.\, e^{{i\over
{\hbar}}\, S_{clas}({\vec x}_{clas}(t), t)}$. This means, however,
that like in the arms of the atom interferometers, in the points of
the screen corresponding to the bright fringes the wave function is
well approximated by an EMWF (modulo corrections of order
$O(\hbar^2)$) corresponding to an \emph{effective} monopole particle
with an \emph{effective} trajectory ${\vec x}_{clas}(t)$ hitting the
screen. Since $S_{class}$ is a stationary action, its variations are
small so that Eq.(\ref{6.7}) implies $< {\hat {\vec p}}_{\psi(t)} >
\approx m\, {\vec v}_B({\vec x}_B(t), t){|}_{{\vec x}_B(t) = {\vec
x}_{clas}(t)}$ and Eqs.(\ref{2.7}) and (\ref{2.8}) imply $ < {\hat
{\vec x}}>_{\psi(t)}\, = {\vec x}_{clas}(t) + O(\hbar^2) \approx
{\vec x}_B(t)$ and $ < {\hat {\vec p}}
>_{\psi(t)} =  {\vec p}_{\psi}(t) = m\, {\dot {\vec x}}_{clas}(t)
\approx m\, {\vec v}_B({\vec x}_B(t) \approx {\vec x}_{clas}(t),
t)$.

\medskip

Let us stress that the non-Newtonian forces of the second order
Newton-like equations of motion of Eq.(\ref{2.9}), acting on an {\it
effective} monopole particle, are not derivable from a potential
(like in general happens with viscous forces). In the chain of
approximations described above, they replace the effect of the
quantum potential in the Hamilton equation (\ref{6.3}) and {\it this
enlightens the difference between the Feynman path} ${\vec
x}_{clas}(t)$ {\it and the Bohmian trajectory} ${\vec x}_B(t)$ {\it
near the points of the bright fringes}.
\medskip

If $\chi^{(bright)}_i$, $-\infty < i < \infty$, are the
characteristic functions identifying the regions of the bright
fringes on the screen and $\chi^{(dark)}_j$, $-\infty < j < \infty$,
the characteristic functions identifying the regions of the dark
fringes, the wave function on the screen is well approximated by an
expression of the type $\psi({\vec x}_{screen}, t) \approx \sum_{i =
-\infty}^{\infty}\, \chi^{(bright)}_i\, \psi_i^{(bright)}({\vec
x}_{screen}, t) + \sum_{j = -\infty}^{\infty}\, \chi^{(dark)}_j\,
\psi_j^{(dark)}({\vec x}_{screen}, t)$, where $\chi^{(bright)}_i\,
\psi_i^{(bright)}({\vec x}_{screen}, t)$ is an EMWF, modulo
corrections of order $O(\hbar)$ and $\chi^{(dark)}_j\,
|\psi_j^{(dark)}({\vec x}_{screen}, t)|^2 \approx O(\hbar^2)$.

\bigskip

We can conclude, therefore, that even in the case of the double
slit, the experiments with a {\it beam} of massive particles could
be interpreted, {\it with all the interference effects taken into
account}, as outcomes of a beam of {\it effective} monopole
particles hitting the screen, rather than as real Bohmian particles.

\bigskip

Let us recall that Bohmian quantum trajectories have recently become
a {\it computational tool} to obtain approximations in many-body
problems connected with molecular dynamics, whenever it is possible
to have a subset of variables of the molecule to be treated quantum
mechanically under the influence of the rest of the system treated
classically. This type of hybrid quantum/classical description
exploits the wave packet of the subsystem to define Bohmian quantum
trajectories which, in turn, are used to calculate the force acting
on the classical variables. See for instance Refs.\cite{36,37,38}
and the associated hydrodynamic formulation of QT \cite{39}. Our
approach suggests that it should be possible to recover the main
results of these approaches by suitably reinterpreting Bohmian
trajectories as {\it effective} trajectories associated with EMWFs.

\vfill\eject

\section{On De-coherence and the Transition from Improper Quantum
Mixtures to Classical Statistical Ones}

\subsection{Preferred Role of Positions and Aspects of Classicality
from De-coherence}

A remarkable contribution to the clarification of the dynamics of QT
and the {\it problem of classical regime} is due to the {\it
de-coherence}-approach. The pervading presence of the environment in
every realistic description of a macroscopic apparatus interacting
with a microscopic system has led to a deeper understanding of the
properties of entanglement. On the one hand, the {\it de-coherence}
approach pointed to the generic existence of a selection of {\it
robust preferred positional bases of pointer states for the
apparatuses} (the only relevant collective variables in the
many-body description of an apparatus that reveal the integer number
resulting from the measurement). On the other hand, it showed the
tendency to an extremely rapid decay of the off-diagonal
interference terms in the reduced density matrix describing the
relevant quantum degrees of freedom of the microscopic system and of
the pointer \footnote{After summation over the environment and the
other many-body degrees of freedom of the apparatus.}.
\medskip

However, it is now commonly agreed that de-coherence is not a
solution to the quantum measurement problem (see, e.g.,
Refs.\cite{1,40,41}). In particular, the reduced density matrix
reflects a {\it quantum improper mixture} mixing 'ignorance and
entanglement' in a non-separable way and not a {\it classical
statistical proper mixture} based on 'ignorance' only. Consequently,
the probabilistic Born rule describes the right experimental results
but cannot be read in terms of classical probability.

\medskip

Anyway, de-coherence supports a preferred role of position
measurements: this fact, together with the assumption we made about
the methodologically autonomous role of space-time {\it vis-a-vis}
quantum theory, once again leads more to a preferred role of the
coordinate representation in Hilbert space.

\subsection{From Improper Quantum Mixtures to Classical Statistical
Ones}

Let us now exploit the results previously obtained in the
two-particle case in order to show that multipolar expansions can be
instrumental in describing the transition from {\it quantum improper
mixtures} to {\it classical statistical ones}.\bigskip

Consider a reduced density matrix ({\it improper mixture}), obtained
for a 'system + apparatus + environment' 3-universe, by means of
de-coherence (no off-diagonal terms after a short time) after having
summed over all the irrelevant variables, except for a pointer
position ${\vec x}_1$ and a particle ${\vec x}_2$

\bea
 \rho^{(red, dec)}_{(o)}({\vec x}_1, {\vec x}_2, t) &=& \sum_w\, p_w\,
 \rho_{(o)(w)}({\vec x}_1, {\vec x}_2, t),  \qquad \sum_w\, p_w =
 1,\nonumber \\
 &&{}\nonumber \\
 \rho_{(o)w}({\vec x}_1, {\vec x}_2, t) &=& < {\vec x}_1, {\vec x}_2
 | {\hat \rho}_{(o)(w)} | {\vec x}_1, {\vec x}_2 > = |\psi_w({\vec x}_1,
 {\vec x}_2, t)|^2,\nonumber \\
 {\hat \rho}_{(o)(w)} &=& |1, 2, w >\,\, < 1, 2, w|,
 \label{7.1}
 \eea

\bigskip

\noindent and perform a multipolar expansion around the two
trajectories ${\vec x}_{1 w}(t)$ and ${\vec x}_{2 w}(t)$, {\it
imposing the vanishing of all the three dipole terms, for each value
of $w$}. Moreover let us assume that all the wave functions be EMWF.

\bigskip

Then, we get

\bigskip

\bea
 <{\hat x^i}_1\, {\hat x^j}_2>_{(red)} &=& \sum_w\, p_w\, <{\hat
 x^i}_1\, {\hat x^j}_2>_{(w)} = \sum_w\, p_w\, \int d^3x_1 d^3x_2 \,
 < 1, 2, w|{\hat x^i}_1\, {\hat x^j}_2\, |1, 2, w
 > =\nonumber \\
 &=&\sum_w\, p_w\, \int d^3x_1 d^3x_2\, x_1^i\, x^j_2\,
 \rho_{(o)(w)}({\vec x}_1, {\vec x}_2, t)
 = \sum_w\, p_w\, x^i_{1 w}(t)\, x^j_{2 w}(t).
 \label{7.2}
 \eea

\bigskip

It is seen therefore that, after de-coherence has suppressed all the
interference terms, the monopole term describes a classical
statistical mixture ({\it ignorance} due to unspecified Cauchy data
inside the volume of the apparatus, with probabilities $p_w$)
obtained from an improper mixture (ignorance + entanglement with
probability $p_w$)

\medskip

The monopole provides a value ${\vec x}_{1 w}(t)$ of the pointer and
${\vec x}_{2 w}(t)$ of the particle position with probability $p_w$,
in a standard measurement. The probabilities are the same as in the
reduced matrix (namely in accord with the Born rule in the framework
of de-coherence). Note that no statement is necessary here about the
collapse of wave functions since it is implied that the measurement
process is ruled by a unitary transformation up to the end. The
underlying strong hypothesis is, of course, that the robust
preferred basis selected by de-coherence must be {\it not only
positional but also such that all the dipoles (and also the {\it
interference terms}) are negligible for every value of $w$}.

\bigskip

In the relativistic case, in order to have the Poincar\'e group
under control, we must consider isolated systems {\it system +
apparatus + environment}, i.e. a 3-universe, where {\it system +
apparatus} is an open subsystem. Another relevant open subsystem is
formed by {\it system + a collective variable of the apparatus
(simulating the pointer)}. In any case, after the separation of the
external non-local non-measurable center of mass described by the
Jacobi data and the elimination of the internal center of mass
inside the Wigner 3-spaces, one is left with a description of the
3-universe in terms of relative variables only. Relativistic
de-coherence must be described in terms of these. The strategy then
is: (i) identifying the {\it important} relative variables to be
measured in order to obtain information on the system (a crucial
relative variable is the one connecting the pointer to the system);
(ii) understanding their dynamics by treating the remaining relative
variables perturbatively \footnote{Like for the Jacobi coordinates
in the description of molecules: choose the Jacobi coordinates
describing the dominating chemical bonds and treat the other
chemical bonds perturbatively.}; (iii) finding the reduced density
matrix depending on the relevant relative variable by summing over
all the other relative variables of the 3-universe; (iv) verifying
whether the wave functions appearing in the reduced density are
EMWFs so that even at the relativistic level one may get a
transition from an improper quantum mixture to a classical
statistical one, for measurements of the relevant relative variable.

\subsection{Beyond De-coherence: Open Quantum Systems}

De-coherence describes relevant properties of {\it a microscopic
system plus a macroscopic apparatus plus environment}, in which only
a collective variable of the apparatus, the pointer, is selected and
described. The difficulty of treating the quantum dynamics of the
many-body degrees of freedom of the macroscopic apparatus is not
really faced, being out of reach, and the final description is often
similar to a standard {\it instantaneous strong impulsive
measurement}. At best, there is an estimate of the time in which the
off-diagonal terms of the reduced density matrix decay.

\medskip

In Ref.\cite{45} there is, instead, a full dynamic treatment of {\it
a spin 1/2 micro-system interacting with an apparatus of $N >> 1$
spins} (the pointer being simulated by a quantum {\it dot} described
by a Curie-Weiss magnet) {\it in equilibrium with a thermal bath of
phonons} (the environment). The total Hamiltonian is explicitly
given. Before the interaction, the pointer is in a meta-stable
paramagnetic state. The Hamiltonian evolution can be studied in
detail and several {\it time scales} emerge in the basis determined
by the interaction Hamiltonian. There is a very short {\it reduction
time} during which the off-diagonal terms of the system-pointer
density matrix decay: re-appearance after a long recurrence time is
avoided by either de-coherence or a dynamic mechanism. The pointer
plus the bath then goes out of equilibrium and the open quantum
macro-system is described via methods similar to the ones of
irreversible quantum statistical mechanics (though with finite $N >>
1$ rather than with the thermodynamic limit) because the interaction
with the micro-system leads to a break of ergodicity. In a so-called
{\it registration time}, the pointer relaxes in a random way to one
of the two possible ferromagnetic states. It should be noted,
however, that the authors of Ref.\cite{45} suggest reading the
unique random outcome of single experiments, but we discarded their
proposal for a solution of the quantum measurement problem in the
framework of a statistical interpretation of QT from the beginning.

\medskip

This example shows that, when one is able to {\it face the dynamics
in some way}, more {\it structure} emerges with respect to the
ordinary de-coherence approaches. See also Ref.\cite{41a} for other
advances in the treatment of open quantum system which seems to be
the main existing technique for dealing with quantum macro-systems.
In any case, an open quantum subsystem generically {\it does not
evolve in a unitary way} \cite{a}.
\medskip

Unfortunately, when the relevant variables of the micro-system and
of the pointer are {\it positional}, we lack analogous dynamic
results in order to deal with the quantum measurement problem.

\vfill\eject

\section{Conclusions}

In this paper we have shown that, given the Hilbert space of the
solutions of the Schroedinger equation in the preferred coordinate
representation for a given quantum system involving massive scalar
particles (either isolated or interacting with an external
potential), it is possible to make a multipolar expansion of the
associated density matrix (and the Wigner function) around a {\it
classical trajectory}. By requiring that the quantum expectation
value of the position operator coincides, at any time, with such
classical trajectory, it follows that a sufficient condition for
this coincidence is that the spatiotemporal multipolar expansion has
a {\it null dipole}. The subset of wave functions giving rise to
expectation values of position and momentum operators satisfying the
Ehrenfest theorem have been named EMWF. The relevance of EMWFs is
the emergence of a notion of {\it effective classical particles} in
the sense that the trajectories chosen for the multipolar expansion
turn out to be solution of second order equations of motion with
forces determined by the quantum Hamiltonian and deviating from
Newtonian (or relativistic) forces by terms in $\hbar^n$  ($n \geq
2)$. Since these wave functions are solutions of the Schroedinger
equation, their {\it whole expansion} contains all the information
about the interference effects present in the given system in any
experiment under study. These classical {\it effective} trajectories
have nothing to do with Bohmian trajectories. Moreover, as highly
desirable, the EMWFs do not satisfy a superposition
principle.\medskip

We have also shown that the extension of these results to N-particle
states allows us to make a transition from a 3N-dimensional
configuration space to N effective classical trajectories in the
Euclidean 3-space for a certain class of wave functions.
\medskip

Our approach suggests the association of such {\it classical
effective particles} with the 'particles' mentioned in the
terminology of experimentalists and phenomenologists when they
describe the preparation, the intermediate steps and the final
outcome of all the actual experiments.

\medskip

Also, the EMWFs provide a direction of solution to an open problem
of de-coherence, since they allow to characterize the transition
from improper quantum mixtures to classical statistical ones in
special cases.

\medskip

Our results have been extended to RQM by using the formulation given
in Ref.\cite{14}.

\medskip

In conclusion, our approach accepts the validity of QT and of its
Born rule, with the associated randomness of the unique outcomes as
a result of measurements, assuming only that the wave function
describes a single quantum system and not an ensemble and that the
mathematical structure of QT is defined in a given spatiotemporal
background. This last assumption implies a preferred role of the
{\it position variables} in accord with the privileged role of
pointer positions, giving us a {\it position signal} as a result of
the unique outcome of the measurements, in the theory of
de-coherence. The same preferred role is present in Bohmian
mechanics in a hidden variable framework. However, our classical
Newtonian trajectories are not Bohmian trajectories and only play
the role of effective emerging physical properties without
ontological status. On the other hand, position variables are very
complex mathematical quantities in QT, as shown in Appendix A. In
recent developments they turn out to be at best {\it unsharp}
observables described by POVM \cite{42,43,44}. The implied {\it bad
localization} is relevant for particle positions (tracks in bubble
chambers, atoms fixed in a cavity,..) but it is only a noise in the
case of macroscopic pointers.

\medskip

The emerging classical trajectories are relevant in supporting
Bohr's point of view for the description of the technically possible
(not merely {\it gedanken}) experiments, but do not imply any
explicit choice among the existing interpretations of the theory of
measurement. Their construction, however, allows to make some
suggestions about the measurement issue if we observe that every
measurement apparatus is a macroscopic quantum object, which should
be described as a {\it many-body system} in a QFT for condensed
matter, and that any measurement lasts a finite period of time
(which can contain different physical time scales for the various
stages of the measurement). Some position collective variable of the
apparatus will describe the pointer. Since the system under
measurement interacts with the apparatus and with a persisting
environment, it is an {\it open subsystem} of the isolated system
which is better described as 'quantum system + quantum apparatus +
quantum environment'. The crucial point of this issue is that the
mutual potentials acting on the subsystems of the global isolated
system cannot in general be simulated as external potential acting
on the subsystem to be investigated and this {\it forbids time
evolution of the latter from being represented by a unitary
transformation}, \cite{a} even if the total isolated system evolves
unitarily. In other words, a break of unitarity at the level of the
subsystem is natural {\it with or without the collapse hypothesis}.
Moreover, in many-body physics, stochastic processes are natural for
both open quantum systems and thermodynamical reasons. All this
means that the standard presentation of quantum systems as unitarily
evolving under the action of an external potential up to measurement
is not a realistic description of the real process but only an
idealization implying a discontinuous break of unitarity in the
evolution of a subsystem in an external potential when the standard
description of an instantaneous measurement is given.

\medskip

Actually, this idealized description of the standard collapse
postulate together with the eigenvalue-eigenvector link makes use of
the almost instantaneous von Neumann measurements, which are a very
poor approximation to the complexity of the above many-body
situation. New types of measurements (weak \cite{c1}, protective
\cite{c2},..) tend to extend the time of measurement. Probably the
main support to the standard point of view indirectly comes from the
success in interpreting atom spectroscopy. The spontaneous
localization of GRW theory \cite{7} is another way to avoid facing
the many-body issue by simulating it with stochastic processes
acting on the investigated system and suitable time scales.
Ref.\cite{45} is the most advanced attempt in treating a many-body
problem in the only area (the spin systems) where we have enough
results to allow to perform explicit calculations. In this way one
gets various time scales for the measurement process which are
compatible with (and even able to extend) the de-coherence results.
\medskip

Our results on the emergence of classical Newtonian trajectories and
on their role in the description of experiments in agreement with
Bohr's point of view suggest that in the idealized framework of an
evolving wave function of a quantum system (the one in which the
postulate of the "wave function collapse" is introduced) the
following scheme should be instrumental. Namely: decomposing the
wave function on a suitable preferred position basis defined near
the surface of the measuring apparatus (i.e., compatible with the
geometry and structure of the detector; maybe in terms of an
over-complete basis like coherent states), as was done in Subsection
B of Section VI for the double slit experiment. In this way the wave
function will appear as a superposition of many EMWFs, whose
associated classical Newtonian trajectories identify different
points on the surface of the apparatus. The suggestion is that, in
each measurement, one of these EMWF components is randomly selected
as the relevant initiator of a many-body amplification process
leading to a unique reading of the pointer consistently with Born's
rule (using the coefficients of the decomposition on the preferred
position basis). Since the readings of the pointer of the detector
are interpretable at the macroscopic level (in Bohr's spirit) as a
classical {\it effective} particle hitting the detector (tracks of
particles, points on the screen of the double-slit experiment,...)
the given suggestion seems reasonable. The other components of the
wave function, not of the EMWF type, will not produce a reading on
the pointer (think of the dark fringes in the double slit
experiment).

\medskip

In conclusion we obtained a kind of two-level structure. A
probabilistic quantum world and a deterministic classical one, in a
mathematically connected and consistent way. With this in view, the
so-called Heisenberg cut between the two worlds becomes an even more
evanescent divide.

A further clarification of this topic will require new experimental
data in the mesoscopic energy region using the best modern
technology and a mathematical effort to find suitable approximations
to the many-body physics relevant to the theory of measurement.

\bigskip

An open problem is the extension of our approach from massive
quantum particles to photons, in the sense of checking whether all
the recent experimental developments in quantum optics allow a
re-interpretation in terms of {\it effective monopole photons}
moving along emergent classical {\it effective rays} of light. In
this case, however, we have no Schroedinger equation and no
Ehrenfest theorem for photons. On the other hand, e.g. in the case
of the double-slit experiment with photons there are {\it weak
measurements of effective average trajectories of single
photons}\cite{21} and proposals of Bohmian trajectories for photons
obtained by studying spin 1 wave equations \cite{62}. Furthermore,
laser beams are visualized as Gaussian beams of classical light at
the macroscopic level and the focalization of photons in the arms of
interferometers can be visualized in terms of light rays of
geometrical optics. Finally, other examples are given by single
atoms trapped in cavities and interacting with laser beams.

\medskip

In order to deal with this problem, one should start from the free
quantum electro-magnetic field in the radiation gauge (where there
is a transverse vector potential ${\vec A}_{\perp}$ representing the
magnetic field and its conjugate momentum, i.e. the transverse
electric field ${\vec E}_{\perp}$) and describe it in the Fock
space. One should then consider the expectation value of the two
quantum fields between coherent states and compare the time
derivatives of these expectation values with the Hamilton equations
of a classical electro-magnetic field in the radiation gauge. Should
an analogue of the Ehrenfest theorem emerge, we could imagine
obtaining emerging Gaussian packets of classical light. Within an
eikonal approximation, they should give rise to rays of light
simulating the trajectories of {\it effective} photons.

Another non trivial open problem is the extension of our approach to
massive particles with spin, described by Pauli and Dirac spinors.

\vfill\eject

\appendix

\section{Mathematical Problems with Localization of Positions}

Our initial assumption was a space-time scenario {\it in which} the
structure of QT must be formulated and discussed, even if its
symbolic structure goes - so to speak - beyond space-time. This
presupposition from the very beginning entails a privileged role of
the position observables and position measurements. We know that
even de-coherence ends up with a privileged role of position
observables. However, the discussion of position measurements has
not come to an agreed upon solution and many problems have not even
reached an accepted formulation yet.

\medskip

In this Appendix we recall the mathematical problems connected with
the localization of massive particles in QT and RQT.

\bigskip

The position measurement for scalar massive particles forces us to
face the problems of {\it unbound position operators with a
continuous spectrum}. It is well-known that this is a source of
highly non trivial problems.\hfill\break

a) In the standard approach, the instantaneous precise measurement
of self-adjoint bounded operators (physical observables) with a
discrete spectrum is formalized by using projection operators (or
projection valued measures, PVM) using the spectral representation
\footnote{If $n$ and $| n >$ are the eigenvalues and eigenvectors of
a self-adjoint bounded operator with discrete non-degenerate
spectrum, the associated PVM is defined by the orthogonal projectors
$P_n = | n >< n |$, $P_n\, P_m = \delta_{nm}\, P_n$, $\sum_n\, P_n =
I_H$, where $I_H$ is the identity operator in Hilbert space. If
$\rho$ is the density matrix of the measured system, the probability
of the outcome $n$ is ${\cal P}_n = Tr (\rho\, P_n)$.}. In this way
one formulates the pragmatic notion of the 'non-unitary collapse' of
the wave function (the projection postulate) and the
eigenvalue-eigenvector link. \medskip

On the other hand, in the case of position measurements we have
unbounded self-adjoint operators and no normalized position
eigenstates (the improper Dirac kets $| \vec x >$, {\it sharp}
eigenstates of the position operator, satisfying $< \vec x | \vec y
> = \delta^3(\vec x - \vec y)$).\medskip

As noted in Ref.\cite{1}, the notion of 'collapse' of the wave
functions requires position wave functions with a finite support
(inside the apparatus). Physically, this is connected to the fact
that an arbitrarily precise measurement of position requires
arbitrarily strong coupling to the system and an arbitrarily large
amount of energy. Therefore one must generalize the standard
formalism to include {\it unsharp} positions with bad localization.
This formalism, also named {\it projection-operator-valued-measures}
(POVM), was introduced in Refs.\cite{42} \footnote{A POVM is a set
of $n$ ($n$ can be bigger than the dimension of the Hilbert space)
self-adjoint non-orthogonal positive semi-definite operators $E_i$
(the 'effects') satisfying $\sum_{i=1}^n\, E_i = I_H$. If $\rho$ is
the density matrix of the measured system, the probability of the
outcome associated with the (non repeatable) measurement of the
operator $E_i$ is ${\cal P}_i = Tr (\rho\, E_i)$. Neumark's dilation
theorem guarantees that every POVM can be implemented as a PVM in a
bigger Hilbert space $H \otimes H_{ancilla}$, where $H_{ancilla}$
can be imagined as describing a measuring apparatus: in this case
one can write $E_i = M_i^{\dagger}\, M_i$ ($M_i$ may be non-positive
operators). Given the $E_i$'s, there are infinite solutions for the
operators $M_i$. This means that the POVM can arise by infinite
different experimental apparatuses (each one described by a
different $H_{ancilla}$).} and then developed in Refs.
\cite{43}.\medskip

The results of measurements of a POVM give imprecise information of
stochastic type on the localization of the particle (see
Refs.\cite{44} for the status of {\it continuous quantum position
weak measurements}\footnote{A {\it continuous weak measurement} is
one in which information is continuously extracted from the system:
one divides time into a sequence of intervals $\triangle\, t$,
considers a weak measurement in each interval, puts the strength of
each measurement proportional to $\triangle\, t$ and finally takes
the limit $\triangle\, t \rightarrow\, 0$. A possible POVM for the
time interval $\triangle\, t$  are the operators $A(\alpha)=
\Big({{4 k \triangle\, t}\over {\pi}}\Big)^{1/4}\,
\int_{-\infty}^{\infty}\, dx\, e^{- 2 k\, \triangle\, t\, (x -
\alpha)^2}\, | x >< x|$ with $\alpha$ a continuous index. If the
initial state is $| \psi > = \int dx\, \psi(x)\, | x >$, the
probability density of measuring the outcome $\alpha$, when
$\triangle\, t$ is small, is $P(\alpha) = Tr[A(\alpha)^{\dagger}\,
A(\alpha)\, | \psi >< \psi |]$. Consequently one gets the following
result for the expectation values of the position $< \alpha > =
\int_{-\infty}^{\infty}\, d\alpha\, \alpha\, P(\alpha) =
\int_{-\infty}^{\infty}\, dx\, x |\psi(x)|^2 = < x >$. A {\it
continuous measurement} results if one makes a sequence of these
measurements and takes the limit $\triangle \rightarrow 0$. In this
way more and more measurements are made in any finite time interval,
but each of them is increasingly weak. Then, a stochastic
Schroedinger equation emerges which describes the evolution of the
original state $| \psi >$ under the effect of the measurements.}.)

\hfill\break

b) The un-sharp POVM of positions analyzed in Ref.\cite{43} have
been recently used to extend the implications of the
Wigner-Araki-Yanase (WAY) theorem\footnote{The total angular
momentum of the object plus apparatus cannot be conserved in an
accurate and repeatable measurement of a particular component.}
\cite{46} (see also Ref.\cite{14}) to unbound self-adjoint
operators. In Ref.\cite{47} there is the analysis of the possibility
of momentum-conserving measurements of the position of a quantum
particle, as well as a discussion of the limitations imposed by the
conservation law to a position measurement of the pointer of an
apparatus. It turns out that, in repeatable measurements, the
position to be measured (either the particle or the pointer) must
commute with the (particle part) of the conserved total momentum. It
is also shown that the WAY theorem precludes accurate and repeatable
measurements of the pointer position (given the momentum
conservation) unless the Yanase condition (i.e., the assumption that
the pointer observable commutes with the conserved quantity) is
fulfilled.\hfill\break

c) As shown in Ref.\cite{48}, the relativistic Newton-Wigner
operator cannot be self-adjoint (again bad localization: macroscopic
tracks of particles in bubble chambers) but only symmetric (see Ref.
\cite{49} for the bad localization implied by symmetric position
operators).\medskip

d) Concerning the experimental and theoretical limits put on the
measurements of position (center of mass) of an atom, see Refs.
\cite{50,51}. In this case, the uncertainty of the localization (of
the order of {\it nanometers}) is much bigger than the Compton
wavelength of the atom itself!

\medskip

Therefore, we still lack a well defined framework to be used in the
area of the foundational problems of QT in all of the approaches,
like de-coherence, where position measurements are crucial.

\vfill\eject


\begin{thebibliography}{}


\bibitem{1}G.Bacciagaluppi, {\it Measurement and Classical Regime in
Quantum Mechanics} (PhilSci-Archive
$http://philsci-archive.pitt.edu/8770/$) to appear in R. Batterman
(ed.), The Oxford Handbook of Philosophy of Physics (Oxford: OUP)..




\bibitem{2}M.Schlosshauer and K.Camilleri, {\it What classicality? Decoherence
and Bohr's Classical Concepts} (arXiv 1009.4072); {\it The
quantum-to-classical transition: Bohr's doctrine of classical
concepts, emergent classicality, and decoherence} (arXiv 0804.1609).

\bibitem{3}M.Schlosshauer, {\it Decoherence, the Measurement Problem
and Interpretations of Quantum Mechanics}, Rev.Mod.Phys. {\bf 76},
1267 (2004).\hfill\break M.Schlosshauer, {\it Decoherence and the
Quantum-to-Classical Transition} (Springer, Berlin,
2007).\hfill\break K.Camilleri, {\it A History of Entanglement:
Decoherence and the Interpretation Problem}, Studies Hist.Phil.
Mod.Phys. {\bf 40}, 290 (2009).






\bibitem{4}N.P.Landsman, {\it Between Classical and Quantum}, in {\it Handbook of the
Philosophy of Science, Vol.2: Philosophy of Physics}, eds.
J.Butterfield and J.Earman, p. 417-553 (Elsevier, Amsterdam, 2006)
(arXiv quant-ph/0506082).


\bibitem{5}H.Janssen, {\it Reconstructing Reality. Enviroment-Induced Decoherence, the
Measurement Problem and the Emergence of Definitness in Quantum
Mechanics. A Critical Assessment}, Master's thesis, theoretical
physics, Redbout University, Nijmegen, 2008
(philsci-archive.pitt.edu/4224/1/scriptie.pdf).

\bibitem{6} M.Genovese, {\it Interpretations of Quantum Mechanics
and the Measurement Problem} (arXiv 1002.0990).

\bibitem{7} G.C.Ghirardi, A.Rimini, and T.Weber, {\it Unified Dynamics for Microscopic
 and Macroscopic Systems}, Phys. Rev. {\bf D34}, 470 (1986).\hfill\break
G.C.Ghirardi, {\it An Attempt at a Macrorealistic Quantum World
View}, in Proceedings of the Symposium {\it The Interpretations of
Quantum Theory: Where Do we Stand ?}, L.Accardi ed., Istituto della
Enciclopedia Italiana, Academy for Advanced Studies USA at Columbia
University, April 1992, Fordham U.P., New York.\hfill\break
G.C.Ghirardi, {\it Collapse Theories}, in {\it Stanford Encyclopedia
of Philosophy, 2002}.

\bibitem{8}D.Bohm and B.J.Hiley, {\it The Undivided Universe}
(Routledge, London, 1993).\hfill\break
 P.R.Holland, {\it The Quantum Theory of Motion} (Cambridge Univ. Press,
 Cambridge, 1993).\hfill\break
 D.D\"urr and S.Teufel, {\it Bohmian Mechanics} (Springer, Berlin, 2009).
 \hfill\break
 O.Passon, {\it Why isn't Every Physicist a Bohmian?}, 2004 (arXiv
 quant-ph/0412119).\hfill\break
 G.Gr\"ubl and S.Kreidl, Bohmian Mechanics group, University of
 Innsbruck, http://bohm-mechanics.uibk.ac.at \hfill\break
 M.Towler, {\it De Broglie-Bohm Pilot-Wave Theory and the
 Foundations of Quantum Mechanics}, graduate lecture course,
 University of Cambridge 2009,
 $http://www.tcm.phy.cam.ac.uk/~mdt26/pilotwaves.htlm$


\bibitem{9}M.Lienert, {\it Pilot Wave Theory and Quantum
Fields},  philsci-archive.pitt.edu/8710/ , Pittsburg.

\bibitem{10}M.Pauri, {\it Epistemic Primacy vs. Ontological
Elusiveness of Spatial Extension: Is There an Evolutionary Role for
the Quantum ?}, Found.Phys. {\bf 41}, 1677-1702 (2011).



\bibitem{11}C.F.v. Weiszs\"acher, {\it The World View of Physics}
(Routledge, London, 1952).

\bibitem{12}N.Bohr, {\it Maxwell and Modern Theoretical Physics},
Nature {\bf 128}, 691 (1931)

\bibitem{13}N.Bohr, {\it Discussions with Einstein on
Epistemological Problems in Atomic Physics}, in {\it Albert
Einstein: Philoser - Scientist}, ed. by P.A.Schlipp, Library of
Living Philosophers, Evanstone, Illinois, 1949, pp.201-241,
reprinted in {\it Quantum Theory and Measurement}, editors
J.A.Wheeler and W.H.Zurek, pp. 9-49 (Princeton University,
Princeton, 1983).

\bibitem{14} A.Peres, {\it Quantum Theory: Concepts and Methods}
(Kluwer, Dordrecht, 1995).\hfill\break
 O.Hay and A.Peres, {\it Quantum and Classical Descriptions of a
Measuring Apparatus}, Phys.Rev. {\bf A58}, 116 (1998) (arXiv
quant-ph/9712044).




\bibitem{15}L.Lusanna, {\it   Canonical Gravity and Relativistic Metrology:
from Clock Synchronization to Dark Matter as a Relativistic Inertial
Effect}, (arXiv 1108.3224); {\it From Clock Synchronization to Dark
Matter as a Relativistic Effect}, to appear in the Proceedings of
the School {\it Black Objects in Supergravity, BOSS2011}, Fracati,
9-13 May 2011 (arXiv 1205.2481).


\bibitem{16}D.Alba, H.W.Crater and L.Lusanna, {\it Relativistic
Quantum Mechanics and Relativistic Entanglement in the Rest-Frame
Instant Form of Dynamics}, J.Math.Phys. {\bf 52}, 062301 (2011)
(arXiv 0907.1816).


\bibitem{17}D.Alba and L.Lusanna,
 {\it Charged Particles and the Electro-Magnetic Field in
Non-Inertial Frames: I. Admissible 3+1 Splittings of Minkowski
Spacetime and the Non-Inertial Rest Frames},  Int.J.Geom.Methods in
Physics {\bf 7}, 33 (2010) (arXiv 0908.0213) and {\it II.
Applications: Rotating Frames, Sagnac Effect, Faraday Rotation,
Wrap-up Effect}, Int.J.Geom.Methods in Physics, {\bf 7}, 185 (2010)
(arXiv 0908.0215). \hfill\break D.Alba and L.Lusanna, {\it
Generalized Radar 4-Coordinates and Equal-Time Cauchy Surfaces for
Arbitrary Accelerated Observers} (2005), Int.J.Mod.Phys. {\bf D16},
1149 (2007) (arXiv gr-qc/0501090).



\bibitem{18}P.Ehrenfest, {\it Bemerking \"uber die angen\"aherte
G\"ultigkeit der klassischen Mechanik innerhalb der
Quantenmechanik}, Z.Phys. {\bf 45}, 455 (1927).


\bibitem{19}G.Friesecke and M.Koppen, {\it On the Ehrenfest Theorem
of Quantum Mechanics}, J.Math.Phys. {\bf 50}, 082102 (2009) (arXiv
0907.1877).\hfill\break
 G.Friesecke and B.Schmidt, {\it A Sharp Version of Ehrenfest's
 Theorem for General Self-Adjoint Operators} (arXiv 1003.3372).

\bibitem{20}R.W.Robinett, {\it Quantum Wave Packet Revivals},
Phys.Rep. {\bf 392}, 1 (2004) (arXiv quant-ph/0401031).

\bibitem{21}W.G.Dixon, {\it Mathisson's New Mechanics: its Aims and
Realizations}, Acta Phys.Pol. Proc.Suppl. {\bf B1}, 27 (2008).

\bibitem{22} Alba D, Lusanna L and Pauri M, {\it New Directions in
Non-Relativistic and Relativistic Rotational and Multipole
Kinematics for N-Body and Continuous Systems}, in {\it Atomic and
Molecular Clusters: New Research}, ed.Y.L.Ping (Nova Science, New
York, 2006) (arXiv hep-th/0505005); {\it Centers of Mass and
Rotational Kinematics for the Relativistic N-Body Problem in the
Rest-Frame Instant Form}, J.Math.Phys. {\bf 43}, 1677-1727 (2002)
(arXiv hep-th/0102087);  {\it Multipolar Expansions for Closed and
Open Systems of Relativistic Particles} , J. Math.Phys. {\bf 46}
062505 (2004) (arXiv hep-th/0402181).


\bibitem{25}S.Kocsis, B.Braverman, S.Ravets, M.J.Stevens,
R.P.Mirin, L.K.Shalm and A.M.Steinberg, {\it Observing the Average
Trajectories of Single Photons in a Two-Slit Interferometer},
Science {\bf 332}, 1170 (2011).



\bibitem{26}W.P.Schleich, {\it Quantum Optics in Phase Space} (Wiley,
Berlin, 2001).\hfill\break
 A.Polkovnikov, {\it Phase Space Representation of Quantum
Dynamics}, Annals of Phys.{\bf 325}, 1790 (2010) (arXiv 0905.3384).

\bibitem{27}L.Lusanna, {\it The N- and 1-Time Classical
Descriptions of N-Body Relativistic Kinematics and the
Electromagnetic Interaction}, Int.J.Mod.Phys. {\bf A12}, 645 (1997).



\bibitem{28}D.Alba, H.W.Crater and L.Lusanna, {\it Hamiltonian
Relativistic Two-Body Problem: Center of Mass and Orbit
Reconstruction}, J.Phys. {\bf A40}, 9585 (2007) (arXiv
gr-qc/0610200).

\bibitem{29}M. Pauri, {\it Invariant localization and mass-spin relations
in the Hamiltonian formulation of classical relativistic dynamics},
Parma University preprint IFPR-T-019, 1971 ~unpublished!; {\it
Canonical (possibly Lagrangian) Realizations of the Poincar\'e Group
with increasing Mass-spin Trajectories}, Invited Talk at the
International Colloquium, Group Theoretical Methods in Physics,
Cocoyoc, Mexico, 1980, Lecture Notes in Physics No. 135, edited by
K. B. Wolf ~Springer-Verlag, Berlin, 1980.

\bibitem{24}M.Dickson, {\it Non-Relativistic Quantum
Mechanics}, in {\it Philosophy of Physics}, Part A, p.275, eds.
J.Butterfield and J.Earman (Elsevier, Amsterdam, 2007).



\bibitem{30}D.Alba, H.W.Crater and L.Lusanna, {\it Towards Relativistic
Atom Physics. I. The Rest-Frame Instant Form of Dynamics and a
Canonical Transformation for a system of Charged Particles plus the
Electro-Magnetic Field and II. Collective and Relative Relativistic
Variables for a System of Charged Particles plus the
Electro-Magnetic Field}, Canad.J.Phys. {\bf 88}, 379 and 425 (2010)
(arXiv 0806.2383 and 0811.0715).




\bibitem{31} H.W.Crater and L.Lusanna, \textit{The Rest-Frame Darwin
Potential from the Lienard-Wiechert Solution in the Radiation
Gauge}, Ann.Phys. (N.Y.) \textbf{289}, 87 (2001).\hfill\break
 D.Alba, H.W.Crater and L.Lusanna, \textit{The Semiclassical
Relativistic Darwin Potential for Spinning Particles in the Rest
Frame Instant Form: Two-Body Bound States with Spin 1/2
Constituents}, Int.J.Mod.Phys. \textbf{A16}, 3365-3478 (2001) (arXiv
hep-th/0103109).\hfill\break
 D.Alba and L.Lusanna, {\it The Lienard-Wiechert Potential of Charged
 Scalar Particles and their Relation to Scalar Electrodynamics in
 the Rest-Frame Instant Form}, Int.J.Mod.Phys. {\bf A13}, 2791
 (1998) (arXiv hep-th/0708156).





\bibitem{32}C.Lammerzahl, {\it The Pseudo-Differential Operator
Square Root of the Klein-Gordon Equation}, J.Math.Phys. {\bf 34},
3918 (1993).

\bibitem{bb} P.Busch, T.Heinonen and P.Lahti, {\it Heisenberg's Uncertainty
Principle}, Phys.Rep. {\bf 452}, 155 (2007).\hfill\break
 P.Busch and P.J.Lahti, {\it On Various Joint Measurements of
 Position and Momentum Observables in Quantum Theory}, Phys.Rev.
 {\bf D29}, 1634 (1984).





\bibitem{33}P.Storey and C.Cohen-Tannoudji, {\it The Feynman Path
Integral Approach to Atomic Interferometry}, J.Phys. II France {\bf
4}, 1999 (1994).\hfill\break
 M.Kasevich and S.Chu, {\it Atomic Interferometry Using Stimulated
 Raman Transitions}, Phys.Rev.Lett. {\bf 67}, 181 (1991).

\bibitem{34}H.M\"uller, A,Peters and S.Chu, {\it Atom
Gravimeters and Gravitational Redshift}, Nature {\it 463}, 926
(2010); {\it Comment on: "Does an Atom Interferometer Test the
Gravitational Redshift at the Compton Frequency?"} (arXiv
1112.6039). \hfill\break
 P.Wolf, L.Blanchet, C.J.Bord\'e, S.Reynaud, C.Salomon and
C.Cohen-Tannoudji, {\it Replay to the Comment on: "Does an Atom
Interferometer Test the Gravitational Redshift at the Compton
Frequency?"} (arXiv 1201.1778); {\it Testing the Gravitational
Redshift with Atomic Gravimeters?} (arXiv 1106.3412);{\it Does an
Atom Interferometer Test the Gravitational Redshift at the Compton
Frequency?}, Class.Quantum Grav. {\bf 28}, 145017 (2011) (arXiv
1012.1194); {\it Atom Gravimeters and Gravitational Redshift},
Nature {\bf 467}, E1 (2010) (arXiv 1009.0602).

\bibitem{35}H.Yabuki, {\it Feynman path integral in the Young double-slit
experiment}, Int.J.Mod.Phys.  25 (1986) 159.

\bibitem{36}E.Gindensperger, C.Meier and J.A.Beswick, {\it Mixing
Quantum and Classical Dynamics using Bohmian Trajectories},
J.Chem.Phys. {\bf 113}, 9369 (2000); {\it Quantum-Classical Dynamics
including Continuum States using Quantum Trajectories}, J.Chem.Phys.
{\bf 116}, 8 (2002).\hfill\break
 B.Poirer, {\it Bohmian Mechanics without Pilot Waves}, Chemical
 Physics, {\bf 370}, 4 (2010).\hfill\break
 D.A.Deckert, D.D\"urr and P.Pickl, {\it Quantum Dynamics with
 Bohmian Trajectories}, J.Phys Chem. {\bf A111}, 10325 (2007).

\bibitem{37}N.Takemoto and A.Becker, {\it Visualization and
Interpretation of Attosecond Electron Dynamics in Laser-Driven
Hydrogen Molecular Ion using Bohmian Trajectories}, J.Chem.Phys.
{\bf 134}, 074309 (2011).

\bibitem{38}CCP6 Workshop on {\it Quantum Trajectories}, eds. K.H.Hughes and
G.Parlant, July 12-14, 2010, Bangor University UK,
$www.ccp6.ac.uk/booklets/CCP6-2011_Quantum_Trajectories.pdf$.

\bibitem{39}C.L.Lopreore and R.E.Wyatt, {\it Quantum Wave Packet
Dynamics with Trajectories}, Phys.Rev.Lett. {\bf 82}, 5190
(1999).\hfill\break
 R.E.Wyatt and E.R.Bittner, {\it Quantum Wave Packet Dynamics with
 Trajectories: Implementation with Adaptive Lagrangian Grids},
 J.Chem.Phys. {\bf 113}, 8898 (2000).\hfill\break
 R.E.Wyatt, {\it Quantum Dynamics with Trajectories: Introduction to
 Quantum Hydrodynamics} (Springer, Berlin, 2005).

\bibitem{40}S.L.Adler, {\it Why Decoherence has not Solved the
Measurement Problem: A Response to P.W.Anderson}, Stud.
Hist.Phil.Mod.Phys. {\bf 34}, 135 (2003).

\bibitem{41}G.Bacciagaluppi, {\it The Role of Decoherence
in Quantum Mechanics}, Stanford Encyclopedia of Philosophy,
ed.E.N.Zalta, 2003 (Fall 2008 edition,
http://plato.stanford.edu/archives/fall2008entries/qm-decoherence/).


\bibitem{45}A.E.Allahverdyan, R.Balian and Theo M. Nieuwenhuizen,
{\it Understanding Quantum Measurement from the Solution of
Dynamical Models} (arXiv  1107.2138).



\bibitem{41a}G.Gualdi and C.P.Koch, {\it Approximating Open Quantum System
Dynamics in a Controlled and Efficient Way: A Microscopic Approach
to Decoherence}, 2011 (arXiv 1111.4959).\hfill\break
 J.Prior, A.W.Chin, S.F.Huelga and M.B.Plenio, {\it Efficient
 Simulation of Strong System-Environment Interactions},
 Phys.Rev.Lett. {\bf 105}, 050404 (2010) (arXiv 1003.5503).

\bibitem{a}A.R.Vargas, {\it Open Quantum Systems and Quantum
Information Theory}, Dissertation thesis at the Ulm University
($http://vts.uni-ulm.de/docs/2011/7581/vts_7581_10841.pdf$).





\bibitem{42} G.Ludwig, {\it Foundation of Quantum Mechanics I and II} (Springer, Berlin,
 1982).\hfill\break
  K.Kraus, {\it States, Effects and Operations: Fundamental Notions
  of Quantum Theory} (Spinger, Berlin, 1983).\hfill\break
 G.M.Prosperi, {\it Theory of Measurement and Second Quantization},
in {\it Mysteries, Puzzles and Paradoxes in Quantum Mechanics}, ed.
R.Bonifacio (American Institute of Physics, New York,  1999).

\bibitem{43}
 P.Busch, P.Lathi and P.Mittelstaedt, {\it The Quantum Theory of
Measurement} (Springer, Berlin, 1996).\hfill\break
 P.Busch, M.Grabowski and P.Lahti, {\it Operational Quantum Physics}, Lecture Notes
in Physics 31 (Springer, Berlin, 1995; 1997 2nd corrected printing).
\hfill\break
 P.Busch, {\it Unsharp Localization and Causality in Relativistic
Quantum Theory}, J.Phys. {\bf A32}, 6535 (1999).

\bibitem{44}C.M.Caves and G.J.Milburn, {\it Quantum-Mechanical Model
for Continuous Position Measurements}, Phys.Rev. {\bf A36}, 5543
(1987).\hfill\break
 K.Jacobs and D.A.Steck, {\it A Straightforward Introduction to
 Continuous Quantum Measurements}, 2006 (arXiv quant-ph/0611067).


\bibitem{c1}S.Parrott, {\it What do Quantum "Weak" Measurements
 Actually Measure?}, 2009 (arXiv 0908.0035).\hfill\break
 D.J.Miller, {\it State-Dependent Rotations of Spins by
Weak Measurements}, Phys.Rev. {\bf A83}, 032121 (2011) (arXiv
 1008.0676).\hfill\break
 J.Drexel, S.Agarwal and A.N.Jordan, {\it Contextual Values of
 Observables in Quantum Meaurements}, Phys.Rev.Lett. {\bf 104},
 240401 (2010).\hfill\break
 P.B.Dixon, D.J.Starling, A.N.Jordan and J.C.Howell, {\it
 Ultrasensitive Beam Deflection Measurement via Interferometric Weak
 Value Amplification}, Phys.Rev.Lett. {\bf 102}, 173601
 (2009) (arXiv  0906.4828).\hfill\break
 G.J.Pride, J.L.O'Brien, A.G.White, T.C.Ralph and H.M.Wiseman, {\it
  Measurement of Quantum Weak Values of Photon Polarization},
  Phys.Rev.Lett. {\bf 94}, 220405 (2005).


\bibitem{c2}Y.Aharonov and L.Vaidman, {\it Measurement of the
Schroedinger Wave of a single Particle}, Phys.Lett. {\bf A178}, 38
(1993).\hfill\break
 Y.Aharonov, J.Anandan and L.Vaidman, {\it Meaning of the Wave
 Function}, Phys.Rev. {\bf A47}, 4616 (1996) and {\it The Meaning of
 Protective Measurements}, Found.Phys. {\bf 26}, 117 (1996).
 \hfill\break
 S.Gao, {\it Interpreting Quantum Mechanics in Terms of
 Random Discontinuous Motion of Particles}, Philosophy Thesis, Sidney
 University 2011, philsci-archive.pitt.edu/8874/ , Pittsburgh.

\bibitem{62}P.Ghose, A.S.Majumdar, S.Ghua and J.Sau, {\it Bohmian Trajectories
for Photons}, Phys.Lett. {\bf A290}, 205  (2001).\hfill\break
 A. S. Sanz, M. Davidovic, M. Bozic and S. Miret-Artes, {\it
Understanding Interference Experiments with Polarized Light through
Photon Trajectories }, Ann.Phys.(NY) {\b 325}, 763 (2010)  (arXiv
0907.2667).

\bibitem{46}E.P.Wigner, {\it  Die Messung quantenmechanischer
Operatoren}, Z.Phys. {\bf 133}, 101 (1952).\hfill\break H.Araki and
M.M.Yanase, {\it Measurement of Quantum Mechanical Operators},
Phys.Rev. {\bf 120}, 622 (1960).\hfill\break
  G.C.Ghirardi, F.Miglietta, A.Rimini and T.Weber, {\it Limitations
  on Quantum Measurements. I. Determination of the minimal Amount of
  Nonideality and Identification of the Optimal Measuring Apparatuses
   and II. Analysis of a Model Example}, Phys.Rev. {\bf
  D24}, 347 and 353 (1981).\hfill\break
S.Luo, {\it Wigner-Yanase Skew Information and Uncertainty
Relations}, Phys.Rev.Lett. {\bf 91 }, 180403   (2003).


\bibitem{47}P.Busch and L.Loveridge, {\it Position Measurements Obeying
Momentum Conservation}, Phys.Rev.Lett. {\bf 106}, 110406 (2011)
(arXiv 1005.0569).\hfill\break
 P.Busch and L.Loveridge, {\it "Measurement of Quantum Mechanical
Operators" Revisited}, to appear in Eur.Phys.J.D (arXiv 1012.4362).




\bibitem{48}J.Butterfield and G.Fleming, {\it Strange Positions}, in {\it
From Physics to Philosophy}, eds. J.Butterfield and C.Pagonis,
pp.108-165 (Cambridge Univ.Press, Cambridge, 1999).


\bibitem{49}A.Kempf, {\it Unsharp Degrees of Freedom and the Generating of
Symmetries}, Phys.Rev. {\bf D63}, 024017 (2000) (arXiv
hep-th/9907160).\hfill\break
 R.T.W. Martin and A.Kempf, {\it Quantum Uncertainty and the Spectra
of Symmetric Operators}, Acta Appl. Math. {\bf 106}, 349
(2009).\hfill\break
 A,Kempf, {\it Spacetime could be Simultaneously Continuous and
Discrete, in the Same Way that Information can be}, New J.Phys. {\bf
12}, 115001 (2010) (arXiv 1010.4354).



\bibitem{50}T.Vaughan, P.Drummond and G.Leuchs, {\it Quantum Limits to
Center-of-Mass Measurements}, Phys.Rev. {\bf A75}, 033617 (2007)
(arXiv cond-mat/0606201).

\bibitem{51}J.Evers, S.Qamar and M.suhail Zubairy, {\it Atom Localization and
Center-of-Mass Wave Functions Determination via Multiple
Simultaneous Quadratura Measurements}, Phys.Rev. {\bf A75}, 053809
(2007)
















\end{thebibliography}
\end{document}